\documentclass[prb,twocolumn,groupedaddress,showpacs,a4paper]{revtex4b5}
\usepackage{graphicx}

\newcommand{\gn}{Gavoret-Nozi\`{e}res }
\newcommand{\zrs}{Zawadowski-Ruvalds-Solana }
\newcommand{\he}{$^4$He }

\newcommand{\dsf}{$S (\vec{q},\omega)$ }

\begin{document}
\bibliographystyle{apsrev}


\title{Theoretical study of the dynamic structure factor
of superfluid \he}



\author{J. Szwabi\'{n}ski}
\author{M. Weyrauch}
\email[]{michael.weyrauch@ptb.de}
\homepage[]{http://www.ptb.de/english/org/q/q1/q102/he4-team.htm}
\affiliation{Physikalisch-Technische Bundesanstalt, D-38116 Braunschweig,
Germany}
\pacs{67.40.-w, 67.40.Db, 61.12.-q}

\date{February 9, 2001}

\begin{abstract}

We study the dynamic structure factor \dsf of superfluid \he
at zero temperature in the roton momentum region and beyond using field-theoretical
Green's function techniques. We start from the Gavoret-Nozi\`{e}res two-particle
propagator and introduce the concept of quasiparticles. We treat the residual (weak)
interaction between quasiparticles as being local in coordinate space and weakly
energy dependent. Our quasiparticle model explicitly incorporates the Bose-Einstein
condensate. A complete formula for the dynamic susceptibility, which is related to $S (\vec{q},\omega)$, is derived. The structure factor is numerically calculated in
a self-consistent way in the special case of a momentum independent interaction between quasiparticles. Results are compared with experiment and other theoretical approaches.

\end{abstract}
\pacs{67.40 Db, 61.12.-q}

\maketitle


\section{Introduction}
\label{Introduction}

The dynamic structure  factor $S(\vec{q},\omega)$ observed in
inelastic neutron scattering experiments on superfluid \he
consists of a sharp peak at the Landau phonon-maxon-roton energy
and a structured continuum at higher energies.  While there is
general agreement that the phonon part of the excitation spectrum
is due to a collective mode of the \he superfluid (zero sound), the
character of the observed sharp structure at higher momentum transfers
has been the subject of considerable debate. This debate is reviewed
in some detail in the books by Griffin~\cite{GRI93} and
Glyde~\cite{GLY94}. In particular, the analysis of the temperature
dependence of the dynamic structure factor suggests that the roton
excitation may be a single particle excitation from the condensate,
while the excitations at intermediate momenta are described as a
mixture of collective and single particle processes
(Glyde-Griffin scenario~\cite{GLY90}).

As was pointed out by Gavoret and Nozi\`{e}res (GN)~\cite{GAV64}, in the
presence of a Bose condensate, the time-ordered density-density
correlation function $\chi^T (\vec{q},\omega)$ of a Bose system splits
into two parts,
\begin{equation}
\label{introduction: chi}
\chi^T (\vec{q},\omega) = \Lambda^{\alpha}(\vec{q},\omega)G_{\alpha}^{\beta}
(\vec{q},\omega)\Lambda_{\beta}(\vec{q},\omega)+\chi_R^T (\vec{q},\omega).
\end{equation}
The second term $\chi_R^T$ describes density-density fluctuations
of non-condensed particles, while the first term explicitly
contains the single particle Green's functions $G_\alpha^\beta$.
This  term is unique to a Bose-condensed liquid and vanishes in
the absence of a condensate. In simple physical terms, one may
interpret this condensate term as a density fluctuation due to
the excitation of   single atoms out of the condensate. According
to the Glyde-Griffin scenario, this interpretation may be useful
in the roton region. For small energies and momenta, however, GN
showed that  both terms in Eq.~(\ref{introduction: chi}) share
the same poles, which are due to compressional sound waves.

Systematic attempts to describe the excitations of superfluid \he
in  the momentum region at and above the roton minimum  within a
phenomenological field-theory began with the work of
Pitaevskii~\cite{PIT59} who showed that the phonon-maxon-roton
curve terminates at twice the roton energy due to quasiparticle
decay. His elegant ideas were developed further by Zawadowski,
Ruvalds and Solana (ZRS)~\cite{ZAW72}. Starting from a
phenomenological Hamiltonian written in terms of quasiparticle
operators, ZRS found a continu\-um in the one-particle spectral
density of states lying above the sharp single-quasiparticle peak.
In contrast to other approaches the work of ZRS
is grounded in a field-theoretical analysis similar to the GN
theory. However, since the density operator in terms of quasiparticles was
not known, ZRS did not calculate the dynamic susceptibility.
Hastings and Halley~\cite{HAS74} attempted to extend the ZRS
theory to calculate the dynamic structure factor by considering
only single quasiparticle contributions to $S(\vec{q},\omega)$.
Their pessimistic view about ZRS-like models may stem from the
incomplete density operators used in their calculations. A more
general expression for the density operator was proposed by
Pistolesi~\cite{PIS98a}, however without a clear connection to the
basic GN theory. As in the ZRS approach,  the role of the Bose
broken symmetry remains unclear, because the condensate does not
appear explicitly within that model.

The goal of our work is three-fold: (a) to find a quasiparticle
model for $S(\vec{q},\omega)$, which is based on the general
formalism of GN and applicable in the roton region and beyond,
(b) to find a connection between this model and the phenomenological
field theory of Pitaevskii and ZRS  and (c) to calculate \dsf consistently
(i.e. under consideration of all terms in Eq.~(\ref{introduction: chi}))
within the model. This entails a consistent calculation of the sharp roton
peak together with the multi-particle continuum.

In Sec.~\ref{theoretical foundations} we discuss the general
formalism of GN introducing our notation and sign convention.
In Sec.~\ref{model}, we introduce the concept of quasiparticles into the
microscopic theory of GN. Treating the interaction between quasiparticles
as being local in coordinate space, we arrive at a model of ZRS-type, but
with an explicit consideration of the Bose condensate and with an expression
for $\chi^T$, which includes both terms in Eq.~(\ref{introduction: chi}). We
study qualitative properties of the model in
Sec. \ref{qualitative considerations}. Thereafter, we present the
iteration scheme used in our numerical calculations of \dsf (Sec.
\ref{technical details}). The results of the calculation are
presented in Sec.~\ref{numerical results} together with comparisons to experiment
as well as to other theoretical calculations. Finally, in Sec.
\ref{conclusions}, we discuss our conclusions and give an outlook for
a further refinement of our calculations.

\section{Theoretical foundation}
\label{theoretical foundations}

In this section we review the GN formalism defining
our notation and sign conventions. As is known from Belaev's work~\cite{BEL58a}, the presence of a Bose condensate introduces `anomalous' propagators. Gavoret and Nozi\`{e}res~\cite{GAV64} systematically extended Beliaev's theory to the two-body propagator. For the structure factor, in particular, this leads to the first term in Eq.~(\ref{introduction: chi}).

\subsection{Response function}
\label{theoretical foundations: response function}

The dynamic structure factor $S(\vec{q},\omega)$, which  is
measured in neutron scattering experiments, is given at zero temperature
by
\begin{equation}
S(\vec{q},\omega)= -{1 \over \pi n}  {\rm Im} \chi^{\rm T}
(\vec{q},\omega) \theta(\omega);\label{response function: s
propto chi}
\end{equation}
the  $\theta$-function cuts off negative frequency contributions
and $n$ denotes the density of the Bose fluid. $\chi^{\rm
T}(\vec{q},\omega)$ is the Fourier transform of the time-ordered
density-density correlation function
\begin{equation}
\label{response function: chiT(omega)}
\chi^{\rm T}(\vec{q},\omega)={1\over 2\pi}\int_{-\infty}^{\infty}
                    {\rm d}t e^{i\omega t} \chi^{\rm T}(\vec{q},t),
\end{equation}
which is defined as
\begin{eqnarray}
\lefteqn{\chi^{\rm T}(\vec{q},t)
                   =  -i \langle {\rm T}\, \rho_{\vec{q}}(t)
                          \rho_{-\vec{q}}(0)   \rangle} \nonumber \\
&&~= -i \sum_{\vec{p},\vec{p}^{\;\prime}} \langle  {\rm T}
                          a^{\dagger}_{\vec{p}-{\vec{q}\over 2}}(t)
                          a_{\vec{p}+{\vec{q}\over 2}}(t)
                          a^{\dagger}_{\vec{p}^{\;\prime}+{\vec{q} \over 2}}(0)
                          a_{\vec{p}^{\;\prime}-{\vec{q}\over 2}}(0)  \rangle .
\label{response function: chiT(t)}
\end{eqnarray}
Here,  $\rho_{\vec{q}}(t) =\sum_{\vec{p}}
a^{\dagger}_{\vec{p}-\vec{q}/2}(t)a_{\vec{p} +\vec{q}/2}(t) $ is
the density operator in the Heisenberg picture and $\rm{T}$ the
time-ordering operator.

\subsection{Two-particle Green's function}
\label{theoretical foundations: two-particle green's function}

Following  Gavoret-Nozi\`eres~\cite{GAV64}, we now consider the
two-particle Green's function,
\begin{eqnarray}
\lefteqn{{K^{\delta \gamma}_{\alpha \beta}
(\vec{p},\vec{p}^{\;\prime},\vec{q};t_{i})}=} \nonumber\\
&&~\langle {\rm T} a^{\delta}_{-\vec{p}^{\;\prime}+{\vec{q}\over 2}}(t_{1})
           a^{\gamma}_{\vec{p}^{\;\prime}+{\vec{q}\over 2}}(t_{2})
           a^{\beta \dagger}_{\vec{p}+{\vec{q}\over 2}}(t_{3})
           a^{\alpha \dagger}_{-\vec{p}+{\vec{q}\over 2}}(t_{4})
  \rangle.
\label{two-particle green's function: K(t)}
\end{eqnarray}
The greek indices take the values 1 or 2 and distinguish between
creation and destruction operators, i.e.
$a^{1}_{\vec{q}}(t)=a_{\vec{q}}(t)$  destructs a Boson (\he atom)
with momentum $\vec{q}$ at time $t$ and $a^{2}_{\vec{q}}(t)
=a^\dagger_{-\vec{q}}(t)$ is a corresponding creation operator.
Greek  subscripts and superscripts on the left hand side of
Eq.~\ref{two-particle green's function: K(t)} label incoming and
outgoing particles, respectively.

It is convenient to work with the Fourier transform of
(\ref{two-particle green's function: K(t)}) with respect to the
time variables:
\begin{eqnarray}
\lefteqn{(2\pi)^4\delta(\omega_3+\omega_4-\omega_1-\omega_2)
K^{\delta \gamma}_{\alpha \beta} (p,p^{\prime},q) =}
\label{two-particle green's function: K(omega)}\\
&&\int \! {\rm d}t_1{\rm d}t_2{\rm d}t_3{\rm d}t_4
e^{i (\omega_4 t_4 +\omega_3 t_3-\omega_2 t_2-\omega_1 t_1)}
K^{\delta \gamma}_{\alpha \beta} (\vec{p},\vec{p}^{\;\prime},\vec{q},t_{i}).\nonumber
\end{eqnarray}
Here, $p=(\vec{p}, \epsilon=(\omega_4-\omega_3)/2)$,
$p^\prime=(\vec{p}^{\;\prime},\epsilon^\prime=(\omega_2-\omega_1)/
2)$ are the  relative momenta and energies of the particle pair
in the initial and final states, respectively, and $q=(\vec{q},
\omega=\omega_3+\omega_4)$ is the total momentum and energy of
that pair.

It should be obvious from Eqs.~(\ref{response function: chiT(t)})
and (\ref{two-particle green's function:
K(t)}) that the dynamic susceptibility can be written as
\begin{equation}
\chi^{\rm T}(q)=-i\sum_{p,p^\prime} K^{21}_{21}(p,p^\prime,q),
\label{two-particle green's function: chi prop to K}
\end{equation}
where we have used the abbreviation $\sum_p \equiv \int \frac{{\rm d}^3p}
{(2\pi)^3}\int \frac{{\rm d}\epsilon}{2\pi}$.
In their diagrammatic analysis of $K_{\alpha \beta}^{\delta
\gamma}$ (see Fig.~\ref{two-particle green's function: diagram of
K}), GN have shown that the dynamic susceptibility may be
separated into two terms, a singular and a regular part,
\begin{figure*}[t]
\includegraphics[height=7cm,width=15cm]{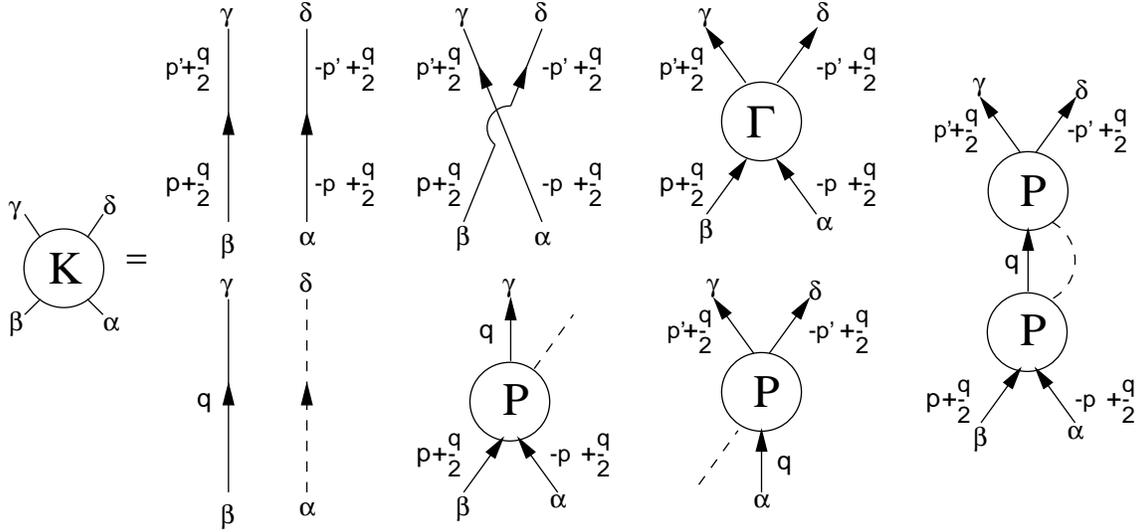}
\caption{Diagrammatic representation of the two-particle Green's
function $K_{\alpha \beta}^{\delta \gamma}$ in energy-momentum
space. The solid lines represent the single-particle Green's functions
$G_{\alpha}^{\beta}$. The dashed lines indicate condensate  particles. ${\bf \Gamma}$
and ${\bf P}$ stand for the four- and three-point vertex function,
respectively. The three diagrams without any condensate line are the
regular two-body  propagator $F_{\alpha \beta}^{\delta \gamma}$ given by
Eq.~(\ref{two-particle green's function: F}). They are the only
contribution to $K_{\alpha \beta}^{\delta \gamma}$ in the absence of
the condensate. All the other diagrams are unique to a Bose-condensed
liquid and lead to the singular part of the dynamic susceptibility
(see. Eq.~(\ref{two-particle green's function: singular part of chi})).}
\label{two-particle green's function: diagram of K}
\end{figure*}
\begin{equation}
\chi^{\rm T}(q) =\chi^{T}_{S}(q) + \chi^{T}_{R}(q).
\label{two-particle green's function: two terms in chi}
\end{equation}

As was already alluded to in the introduction,
the singular susceptibility  $\chi^{T}_{S}(q)$ arises due to the
Bose condensate and is given by
\begin{equation}
\chi^{T}_{S}(q) = \Lambda^{\alpha}(q)
                           G_{\alpha}^{ \beta}(q)
                           \Lambda_{\beta}(q).
\label{two-particle green's function: singular part of chi}
\end{equation}
In Eq.~(\ref{two-particle green's function: singular part of chi})
and elsewhere in this paper,  summation over repeated indices is
implied. The singular part of $\chi^T$ contains the Beliaev single particle
Green's functions~\cite{BEL58a} $G_{\alpha}^{ \beta}(q)$. Consequently,
any structure in $G_{\alpha}^{ \beta}(q)$ will show up directly in
the dynamic susceptibility. This is an interesting physical
feature of a partly condensed Bose liquid: excitations of single
particles out of the condensate determine partly the density
fluctuations of the liquid. The Beliaev Green's functions obey
the Dyson-Beliaev equation,
\begin{equation}
 G_\alpha^\beta(q)= {G_0}_\alpha^\beta(q)+{G_0}_\alpha^\eta(q)
{\Sigma}_\eta^\xi(q){G}_\xi^\beta(q)
\label{two-particle green's function: beliaev-dyson equation}
\end{equation}
with the free Green's functions
\begin{eqnarray}
{G_0}_1^1(q) & = & {G_0}_2^2(-q)=(\omega-\epsilon_{\vec{q}}
                   +\mu+i\eta)^{-1}, \nonumber \\
{G_0}_1^2(q) & = & {G_0}_2^1(q) = 0.
  \label{two-particle green's function: G0}
\end{eqnarray}
and the self energies ${\Sigma_\eta^\xi}(q)$'s. The free-atom
kinetic energy is denoted by $\epsilon_{\vec q}$ and the chemical
potential by $\mu$. The Bose vertex functions
$\Lambda^{\alpha}(q)$ determine how strongly single particle
excitations contribute to the dynamic susceptibility,
\begin{equation}
\label{two-particle green's function: lambda}
\Lambda^{\alpha}(q)  =  n_{0}^{1/2} ( \delta_{2 \alpha}+\delta_{1 \alpha})
                          +i\sum_p {F_0}_{21}^{\zeta\epsilon}(p,q)
                          P^{\alpha}_{\zeta \epsilon}(p,q).
\end{equation}
Here, $n_0$ denotes the density of condensed bosons. These
functions vanish in the absence of the condensate.  The
interaction vertex (full three point function) ${\bf P}$ will be
discussed in more detail below. Here and elsewhere in this paper,
a boldfaced symbol stands for matrix functions. The product of two
Belaev single particle propagators ${\bf G}$ with total momentum
$q$ is abbreviated as
\begin{equation}
{F_0}_{\alpha\beta}^{\gamma\delta}(p,q)=G_{\alpha}^{\delta}(-p+{q\over 2})
                       G_{\beta}^{\gamma}(p+{q\over 2}).
\label{two-particle green's function: F0}
\end{equation}

The regular susceptibility $\chi^{T}_{R}$ is given by
\begin{equation}
\chi^{T}_{R}(q)=-i\sum_{p,p^\prime}F^{21}_{21}(p,p^{\prime},q),
\end{equation}
in terms of the regular two-body propagator
\begin{widetext}
\begin{equation}
F_{\alpha \beta}
^{\delta \gamma}
(p,p^{\prime},q)  =  -(2\pi)^4
                  \left[{F_0}_{\alpha\beta}^{\delta\gamma}(p,q)\delta (p-p^{\prime})+
                   {F_0}_{\beta\alpha}^{\delta\gamma}(p,q)\delta (p+p^{\prime})\right]
                    -i{F_0}_{\alpha\beta}^{\zeta\epsilon}(p,q)
                    \Gamma^{\eta \xi}_{\zeta \epsilon}(p,p^\prime,q)
                    {F_0}_{\eta\xi}^{\delta\gamma}(p^\prime,q).
\label{two-particle green's function: F}
\end{equation}
\end{widetext}
The regular two-body  propagator is determined by the single
particle propagator ${\bf G}$ and the (full) two-body interaction
kernel $\bf{\Gamma}$. Terms where the condensate does not
contribute will be called `regular' in this paper. $\chi^{T}_{R}$
represents the full  response function of a Bose liquid in
the absence of the condensate.

Following GN we separate out processes with intermediate two particle
states in the interaction kernel $\bf{\Gamma}$, so
that it fulfills the following (coupled) Bethe-Salpeter equations,
\begin{eqnarray}
\lefteqn{\Gamma^{\eta \xi}_{\zeta \epsilon} (p,p^{\prime},q)
    =I^{\eta \xi}_{\zeta \epsilon} (p,p^{\prime},q)}
     \label{two-particle green's function: Gamma}\\*
&&~~+{1\over 2}i\sum_{p
                                   ^{\prime \prime}}
                  I^{\mu \nu}_{\zeta \epsilon}(p,p^{\prime \prime},q)
                  {F_0}_{\mu\nu}^{\tau\lambda}(p^{\prime\prime},q)
                  \Gamma^{\eta \xi}_{\tau \lambda}(p^{\prime
                                    \prime},p^{\prime},q).\nonumber
\end{eqnarray}
$\bf{I}$ is the  two-particle irreducible interaction kernel and
contains all diagrams without intermediate two-particle states.
The full interaction kernel $\bf{\Gamma}$ not only determines the
structure of the regular susceptibility but also the structure of
the three-point kernel $\bf{P}$ and, consequently, the structure
of the self energy $\bf{\Sigma}$. Using graphical methods one can
show, that, in terms of $\bf{\Gamma}$, the three point function
$\bf{P}$ is given by
\begin{eqnarray}
\lefteqn{P^{\alpha}_{\zeta \epsilon}(p,q)=
   J^{\alpha}_{\zeta \epsilon}(p,q)}\nonumber\\*
&&~~+{1\over 2}i\sum_{p^\prime}
                  \Gamma^{\mu \nu}_{\zeta \epsilon}(p,p^{\prime},q)
                  {F_0}_{\mu\nu}^{\eta\xi}(p^\prime,q)
                  J^\alpha_{\eta \xi}(p^{\prime},q),
\label{two-particle green's function: P}
\end{eqnarray}
and the self energy by
\begin{equation}
{\Sigma}_\alpha^\beta(q)={\tilde{\Sigma}}_\alpha^\beta(q)+
{{\Sigma^*}}_\alpha^\beta(q)
\label{two-particle green's function: Sigma}
\end{equation}
with
\begin{equation}
{ \Sigma^*}_\alpha^\beta(q)={1\over 2}i\sum_p
                { J}^{\zeta\epsilon}_\alpha(p,q)
                {F_0}^{\mu\nu}_{\zeta\epsilon}(p,q)
                { P}^\beta_{\mu \nu}(p,q).
\label{two-particle green's function: Sigma*}
\end{equation}
$\bf{J}$ and $\bf{\tilde{\Sigma}}$ are again two-particle
irreducible functions and contain all diagrams without
intermediate two-particle states. As is obvious from these
equations, the various terms $\bf{G,P}$, and $\bf{\Gamma}$, which
determine the structure of the susceptibility, are not
independent, but are linked in a complicated way. For instance,
any singularity (i.e. structure)  in $\bf{\Gamma}$ will show up
in all terms, and care must be taken to treat them consistently.

In order to put the general formalism outlined above to work, the
two-particle irreducible functions $\bf{\tilde{\Sigma}}$, $\bf{J}$
and $\bf{I}$ must be determined. GN have done this in the hydrodynamic
limit $\vec{q},\omega \rightarrow 0$. But in general, this is a most
formidable task. Therefore, the formalism is only useful if $\bf{\tilde{
\Sigma}}$, $\bf{J}$ and $\bf{I}$ can be approximated in a simple way.

\section{Quasiparticle Model}
\label{model}

In the following we will introduce the approximations
appropriate for the momentum regime we are interested in. We are guided by the
the pioneering work of Pitaevski~\cite{PIT59} as well as the ZRS
phenomenological field theory~\cite{ZAW72}.

\subsection{Quasiparticles}
\label{model: quasiparticles}

Starting from Eq.~(\ref{two-particle green's function: Sigma})
we transform the Beliaev-Dyson matrix equation (\ref{two-particle
green's function: beliaev-dyson equation}) into the form
\begin{equation}
G_\alpha^\beta(q)  =  {g}_\alpha^\beta(q)+{g}_\alpha^\eta(q)
                         {\Sigma}_\eta^{*\xi}(q){G}_\xi^\beta(q)
\label{quasiparticles: beliaev-dyson equation}
\end{equation}
with
\begin{equation}
 g_\alpha^\beta(q)  =  {G_0}_\alpha^\beta(q)+{G_0}_\alpha^\eta(q)
                         \tilde{\Sigma}_\eta^{\xi}(q){g}_\xi^\beta(q).
\label{quasiparticles: beliaev-dyson equation for g}
\end{equation}
At this point it is convenient to introduce the concept of
quasiparticles. Eq.~(\ref{quasiparticles: beliaev-dyson
equation}) is  nothing but an expression for Green's
functions describing the propagation of particles with `bare'
propagators given by the $g_\alpha^\beta(q)$'s. Thus, we assert that,
in fact, $g_\alpha^\beta(q)$ describe  stable
quasiparticles, which are helium atoms renormalized by the
two-particle reducible part of the self energy. If this
interpretation is at all useful, then the residual interaction of
the quasiparticles must be weak. We will exploit this
interpretation in the following: All the strong atomic
interactions are contained in ${\bf \tilde\Sigma}$ generating the
`bare' quasiparticle spectrum, which, of course, is not at all
parabolic as for a free He atom.

Eq.~(\ref{quasiparticles: beliaev-dyson equation}) with the self energy
(\ref{two-particle green's function: Sigma*}) precisely
corresponds to the Dyson equation used by Pitaevskii \cite{PIT59}
as the basis of his evaluation of quasiparticle spectrum near
its  endpoint. Moreover, the ZRS phenomenological field theory is
built on a similar basis. Thus one can identify Pitaevskii-ZRS
quasiparticles with helium atoms renormalized by ${\bf
\tilde{\Sigma}}(q)$.

In order to further specify  the quasiparticle we have
to define the ${\tilde\Sigma}_{\alpha}^{\beta}$ or alternatively
$g_\alpha^\beta$. In this work, we will take

\begin{equation}
g_\alpha^\beta(q)  =  \frac{A_\alpha^\beta({\vec q})}
                     {\omega-\omega_{{\vec q}}^0+i\eta}.
\label{quasiparticles: g}
\end{equation}
This kind of propagator characterizes a stable quasiparticle with
the `bare' spectrum $\omega_{\vec{q}}^0$ and normalization factor (residue)
$A_{\alpha}^{\beta}(\vec{q})$. In general, such a Green's
function should have a second pole at negative energies, but it
contributes only a small correction at high energies (if
$kT\ll \omega_{\vec{q}}^0$) and therefore, it can be neglected in our
context.

Furthermore, we assume that the residues $A_{\alpha}^{\beta}$ fulfill the relation
\begin{equation}
A_1^1 (\vec{q})A_2^2 (\vec{q}) - A_2^1 (\vec{q})A_1^2 (\vec{q})=0.
\label{quasiparticles: residues}
\end{equation}
The same relation holds in the Bogoliubov model~\cite{BOG47}
of a dilute weakly interacting Bose gas. It
simplifies a formal algebraic
solution of the Beliaev-Dyson equation (\ref{quasiparticles:
beliaev-dyson equation}) significantly and leads to a ZRS-like
expression for $G_{\alpha}^{\beta}(q)$,
\begin{equation}
\label{quasiparticles: G}
G_{\alpha}^{\beta}(q)={A_{\alpha}^{\beta}(\vec{q}) \over \omega -
\omega_{\vec{q}}^0 - A\Sigma^*(q)}
\end{equation}
with
\begin{equation}
\label{quasiparticles: asigma}
A\Sigma^*(q)=A_{\mu}^{\nu}(\vec{q}){\Sigma^*}_{\nu}^{\mu}(q).
\end{equation}
In order to make contact with the  original ZRS expressions, which
contain only one Green's function, we have to make some specific
choices for the renormalization functions $A_{\alpha}^{\beta}$ in
Eq.~(\ref{quasiparticles: G}). Typically, one assumes that the ZRS
calculation scheme corresponds to keeping only the diagonal
element $G_1^1$ of the Beliaev-Green's matrix (see e.g. Chap. 10
of Ref.~\onlinecite{GRI93}) and takes $A_1^2=A_2^1=A_2^2=0$ and $A_1^1=1$.
This choice is also inspired by the Bogoliubov approximation. One
may argue that at large wavevectors the anomalous propagators are
much smaller than the diagonal ones because
the Bose coherence factors become unimportant as in the Bogoliubov
model. However, since we do not know the explicit form of these
factors in our model, we will keep the anomalous Green's
functions  in the following. At the beginning of the
Sec.~\ref{numerical results}, we will present an alternative
choice of the parameters $A_{\alpha}^{\beta}$, which is also
consistent with ZRS scheme.

\begin{figure}[t]
\includegraphics[height=60mm,width=70mm]{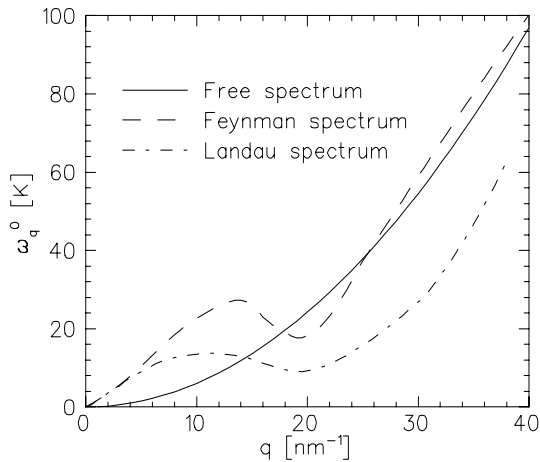}
\caption{Two possible `bare' quasiparticle spectra: the Landau
spectrum (dashed-dotted line) and the Feynman spectrum (dashed
line) compared with the free \he energy $q^2/2m$ (solid line).}
\label{quasiparticles: free quasiparticle spectra}
\end{figure}

Finally, we suppose that the `bare' quasiparticle spectrum is
already characterized by a roton minimum and the residual
interaction does not modify the spectrum qualitatively in the
momentum region up to the roton minimum. This assumption is in
the spirit of our approach assuming weakly interacting quasiparticles.
Since  an unambiguous determination of the
`bare' spectrum is impossible, we will study two models with different
spectra (see Fig.~\ref{quasiparticles: free quasiparticle spectra}):
(i) Landau spectrum~\cite{LAN47,ZAW72}, which was originally
    chosen in order to explain the specific heat data and
(ii) Bijl-Feynman spectrum~\cite{FEY54}, proposed first by
             Bijl and then derived by Feynman.
Both spectra  are qualitatively similar, but differ in
the roton and maxon energies.

\subsection{Response function}
\label{model: response function}

If the concept of quasiparticles is
at all useful, the residual interaction of the quasiparticles
must be weak. Thus, the interaction energy in the momentum region
of interest should be much smaller than the kinetic energy of
quasiparticles and the interaction vertices can be treated as
being essentially local in coordinate space and weakly energy
dependent. The Fourier transform of a local interaction vertex is
a function of only the total momentum transfer, i.e.
\begin{equation}
{\bf J}(p,q)\simeq{\bf J}({\vec q}), \;\;\;\;\;
{\bf I}(p,p^\prime,q)\simeq{\bf I}({\vec q}).
\label{response function (model): interaction}
\end{equation}
As we shall see, this corresponds to an RPA-like approximation.

In order to determine the dynamic susceptibility in line with
Eq.~(\ref{response function (model): interaction}) we first define
\begin{equation}
{f_0}^{\zeta\epsilon}_{\mu\eta}(q)=i\sum_p
{F_0}_{\mu\nu}^{\zeta\epsilon}(p,q); \;\;\;
{f}^{\zeta\epsilon}_{\mu\eta}(q)=-i\sum_{p,p^\prime}
{F}_{\mu\nu}^{\zeta\epsilon}(p,p^\prime,q).
\label{response function (model): f and f0}
\end{equation}
From Eq.~(\ref{two-particle green's function: F}) one finds
\begin{equation}
f_{\alpha\beta}^{\delta\gamma}(q)={f_0}_{\alpha\beta}^{\delta\gamma}(q)
                                +{f_0}_{\beta\alpha}^{\delta\gamma}(q)
                                +{f_0}_{\beta\alpha}^{\mu\nu}(q)
                                 \Gamma_{\mu\nu}^{\eta\xi}(\vec{q})
                                 {f_0}_{\eta\xi}^{\delta\gamma}(q).
\label{response function (model): f}
\end{equation}
Using  Eqs. (\ref{response function (model): interaction}), (\ref{response
function (model): f and f0}) and (\ref{response function (model): f}) in conjunction with Eq.~(\ref{two-particle green's function: P}) we can express the reducible part of the self
energy~(\ref{two-particle green's function: Sigma*})  in terms of the  quantity
$f_{\alpha\beta}^{\delta\gamma}(q)$,
\begin{equation}
{ \Sigma^*}_\alpha^\beta(q)={1\over 4}
                { J}^{\mu\nu}_\alpha({\vec q})
                {f}^{\xi\eta}_{\mu\nu}(q)
                { J}^\beta_{\xi \eta}({\vec q}).
\label{response function (model): Sigma*}
\end{equation}

The separation of the dynamic susceptibility (\ref{two-particle
green's function: two terms in chi}) into a singular and a regular
part is very useful, since it emphasizes the role of the
Bose-Einstein condensate, which couples  the single-quasiparticle
propagator ${\bf G}$ into the
density-density correlation function $\chi^{\rm T}(q)$. However,
for practical calculations it is sometimes more convenient to
sort the diagrams for $K^{\delta \gamma}_{\alpha \beta}
(p,p^{\prime},q)$ differently: Let $W^{\delta
\gamma}_{\alpha \beta} (p,p^{\prime},q)$ be made up of all diagrams
with  momenta of the external lines different from zero (see
Fig.~\ref{response function (model): diagram of W}),
\begin{figure}[!t]
\includegraphics[height=4.9cm,width=5.4cm]{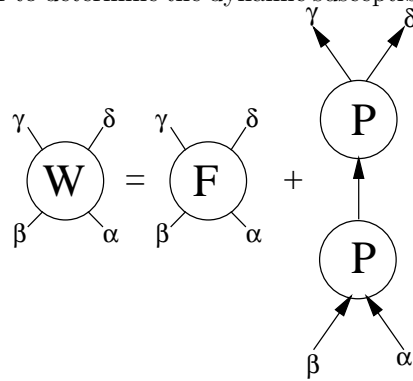}
\caption{Diagrammatic representation of the two-particle
propagator ${\bf W}$ (see Eq.~(\ref{response function (model): W})) with all external momenta different from zero.
${\bf F}$ is the  regular two-particle Green's function given by
Eq.~(\ref{two-particle green's function: F}).} \label{response
function (model): diagram of W}
\end{figure}
\begin{widetext}
\begin{equation}
W^{\delta \gamma}_{\alpha \beta}(p,p^{\prime},q)
  =  F^{\delta \gamma}_{\alpha \beta} (p,p^{\prime},q)
-iG_{\alpha}^{\zeta}(-p+{q\over  2})G_{\beta}^{\epsilon}(p+{q\over 2})
P_{\zeta \epsilon}^{\rho}(p,q)G_{\rho}^{\sigma}(q)P_{\sigma}^{\eta \xi}(p^{\prime},q)
G_{\eta}^{\delta}(-p^{\prime}+{q\over 2})G_{\xi}^{\gamma}(p^{\prime}+{q\over 2});
\label{response function (model): W}
\end{equation}
\end{widetext}
${\bf W}$ precisely corresponds  to the two-particle Green's
functions introduced by Fukushima and Iseki~\cite{FUK88}.
Apart from the regular two-particle propagator ${\bf F}$ it
contains a term, which  renormalizes  ${\bf F}$ by taking into
account the possibility that two quasiparticles interact via the three
point vertex ${\bf P}$ and propagate as one quasiparticle in an
intermediate step. It follows from Eq.~(\ref{two-particle green's
function: lambda}) that this second term belongs to $\chi^{T}_{S}(q)$
in Eq.~(\ref{two-particle green's function: two terms in chi}).

With Eq.~(\ref{response function (model): W}) we can write the
dynamic susceptibility in the form
\begin{eqnarray}
\lefteqn{\chi^T(q)=n_0\sum_{\alpha\beta}G_\alpha^\beta(q)}
\label{response function (model): chi}\\*
&&~+n_0^{1/2}\left[G_2^\beta(q)+G_1^\beta(q)\right]
J_\beta^{\eta\xi}({\vec q})f^{21}_{\eta\xi}(q)+w^{21}_{21}(q)\nonumber,
\end{eqnarray}
where the function $w_{21}^{21}(q)$ is given by
\begin{eqnarray}
\lefteqn{w_{21}^{21}(q)\equiv   -i\sum_{p p^{\prime}}W^{21}_{21}
                  (p,p^{\prime},q)} \label{response function (model): w} \\*
&&~~=  f_{21}^{21}(q)+{f_0}^{\zeta\epsilon}_{21}(q)
                  P_{\zeta \epsilon}^{\rho}(q)G_{\rho}^{\sigma}(q)
                  P_{\sigma}^{\eta \xi}(q){f_0}^{21}_{\eta \xi}(q).\nonumber
\end{eqnarray}
A brief discussion of expression (\ref{response function (model):
chi}) is in order: The first term describes one-quasiparticle
excitations and vanishes  in the absence  of the condensate. The
third term corresponds to the direct excitation of two
quasiparticles. Since in the absence of the condensate $W_{\alpha
\beta} ^{\delta \gamma}$ reduces to the regular two-particle
propagator, this term goes over into the full response function
above $T_\lambda$. The second term describes an interference
between the one- and two-particle channel and it disappears as
well in the absence of the condensate. Eq.~(\ref{response
function (model): chi}) shows how these three terms must be
combined in order to calculate the susceptibility. In earlier
literature, the dynamic structure factor was either calculated
from the imaginary part of the single particle propagator  (first term in Eq.~(\ref{response function (model): chi}), see e.g.
Ref. \onlinecite{HAS74}) or from the imaginary part of the two-body
propagator $w_{21}^{21}$ (e.g. Ref. \onlinecite{FUK88}). The interference term was
neglected. Juge and Griffin~\cite{JUG94} emphasized that all
terms  may be important,
however they did not combine them in order to calculate
$S(\vec{q},\omega)$.

It is possible to express the susceptibility entirely in terms of
$g_{\alpha}^{\beta}(q)$ and $f_{\alpha \beta}^{\delta
\gamma}(q)$. To this end, we transform first the expressions
(\ref{response function (model): f}) and (\ref{response function
(model): w}) by using of (\ref{two-particle green's function:
Gamma}) and~(\ref{two-particle green's function: P}),
\begin{eqnarray}
\lefteqn{f_{\alpha \beta}^{\delta \gamma}(q)
    = {f_0}_{\alpha \beta}^{\delta \gamma}(q)}\nonumber\\*
&&~~+{f_0}_{\beta\alpha}^{\delta\gamma}(q)
    +\frac{1}{2}{f_0}_{\alpha\beta}^{\mu\nu}(q)
    I_{\mu\nu}^{\eta\xi}(\vec{q})f_{\eta\xi}^{\delta\gamma}(q),
    \label{response function (model): equation for f}\\
\lefteqn{w_{21}^{21}(q) = f_{21}^{21}(q)}\nonumber\\*
&&~~+\frac{1}{4}f_{21}^{\zeta\epsilon}(q)
                     J^\mu_{\zeta\epsilon}({\vec q})g_\mu^\nu(q)
                     J_\nu^{\eta\xi}({\vec q})
                     w_{\eta\xi}^{21}(q).
\label{response function (model): equation for w}
\end{eqnarray}
Obviously, we have found a set of algebraic equations for
$f_{\alpha \beta}^{\delta \gamma}(q)$ and $w_{21}^{21}(q)$, which
are similar to the Dyson-Beliaev equation
(\ref{quasiparticles: beliaev-dyson equation}) for
$G_{\alpha}^{\beta}(q)$.  We can solve them formally to obtain
\begin{widetext}
\begin{eqnarray}
\chi^T (q) & = & n_0 \sum_{\alpha\beta} \left[({\bf 1}- \frac{1}{4} {\bf g}(q)
             {\bf J}^T(\vec{q}){\bf f}(q){\bf J}(\vec{q}))^{-1}
             \right]_\alpha^\rho g_\rho^\beta(q)+
             n_0^{1/2} \sum_\alpha \left[({\bf 1}- \frac{1}{4} {\bf g}(q)
             {\bf J}^T(\vec{q}){\bf f}(q){\bf J}(\vec{q}))^{-1}
             \right]_\alpha^\rho g_\rho^\beta(q)
             J_\beta^{\eta\xi}(\vec{q})f_{\eta\xi}^{21}(q) \nonumber \\
       &   & +\left[({\bf 1} - \frac{1}{4} {\bf f}(q)
             {\bf J}(\vec{q}){\bf g}(q){\bf J}^T(\vec{q})^{-1}
             \right]_{21}^{\alpha\beta}f^{21}_{\alpha\beta}(q)
\label{response function (model): chi in RPA form}
\end{eqnarray}
\end{widetext}
with $f_{\rho \delta}^{\delta \gamma}(q)$ given in terms of
${f_0}_{\alpha \beta}^{\delta \gamma}(q)$
\begin{equation}
f^{\delta\gamma}_{\rho\sigma}(q)=\left[({\bf 1} - \frac{1}{2}{\bf
f_0}(q)
 {\bf
I}(\vec{q})^{-1}\right]^{\alpha\beta}_{\rho\sigma}({f_0}^{\delta\gamma}_{\alpha\beta}
(q)+{f_0}^{\delta\gamma}_{\beta\alpha}(q)); \label{response
function (model): f in terms of f0}
\end{equation}
obvious matrix notation is employed. The denominators in
(\ref{response function (model): chi in RPA form}) clearly show
the RPA character of the approximation applied here. Note, that
all terms in $\chi^{\rm T}(q)$ have the same pole structure. Thus
both the sharp peak and the continuum observed in
$S(\vec{q},\omega)$ receive contributions from all three terms. It
may be useful to stress, that up to this point, no approximations
are involved in the derivation of (\ref{response function
(model): chi in RPA form}) apart from Eq.~(\ref{response function
(model): interaction}). However, in order to explicitly evaluate $\chi^T$
we will need the approximations for $g_{\alpha}^{\beta}$ made in
Sec.~\ref{model: quasiparticles}.

We want to end this section  by noting that an expression similar
to Eq.~(\ref{response function (model): chi}) was proposed by
Pistolesi \cite{PIS98a} as an extension of the \zrs work. His
Eq.~(4) corresponds term by term to our Eq.~(\ref{response
function (model): chi}). But there are also
some differences between both calculations. First of all, the
Bose-Einstein condensate does not appear explicitly in
Pistolesi's expression. Instead  he introduces two
fitting parameters. Secondly, he deals with only one Green's
function and the connection to the general formalism of \gn
remains unclear. To evaluate $\chi^T$ Pistolesi extracts the regular
part of the two-quasiparticle propagator
from experimental data. In our calculation (see
Sec.~\ref{technical details}), we will calculate it within
an iteration scheme.

\section{Qualitative considerations}
\label{qualitative considerations}

In this section, we will qualitatively study two specific kinematical
regions, which deserve special attention: the endpoint region and
quasifree scattering at high energy and momentum transfers.

\subsection{Endpoint of the spectrum}
\label{qualitative considerations: endpoint}

As has been shown by Pitaevskii~\cite{PIT59}, the  renormalized
single-quasiparticle spectrum of a Bose liquid has an endpoint,
i.e. at zero temperature undamped excitations cannot
exist at momenta larger than some threshold value, provided the
`bare' quasiparticle spectrum is characterized by a minimum.
Pitaevskii has clarified the character of the spectrum near its
endpoint in a quite general way by analyzing the singularities of
the single-quasiparticle propagator.

Pitaevskii started from the following expression for the self
energy (written in our notation):
\begin{equation}
\label{endpoint: Sigma* (pit)}
\Sigma^*(q)={i\over (2\pi)^4} \int {\rm d}^4 p P(p,q)G(p)G(q-p)J(p,q)
\end{equation}
which is similar to our Eq.~(\ref{two-particle green's
function: Sigma*}). The only difference is that Pitaevskii did
not consider all Beliaev-Green functions as we do. However, since
all these functions have the same pole structure (see e.g.
Ref.~\onlinecite{BEL58a}), near the singularities our expression should
lead to the same results except for some coefficients and/or
regular contributions, which are not of interest for the analytical
behavior of the propagators. Thus Pitaevskii's
prediction about the endpoint of the renormalized spectrum due to
the decay of quasiparticles holds within our model.

Pitaevskii distinguished  three kinds of decay processes into two
excitations: (i) phonon creation, (ii) decay into excitations with
finite momenta propagating in the same direction with the same
velocity and (iii) decay into two rotons.

Since we are only interested in the momentum region $10 \leq
\vec{q} \leq 40$ nm$^{-1}$ and do not take phonons into account,
the emission of two rotons is the only
possible decay channel. Thus, the spectrum should
have an endpoint of the third kind, which is characterized by a
logarithmic singularity. This endpoint leads to a threshold at
twice the roton energy in the imaginary part of the self energy,
i.e. to
\begin{equation}
\label{endpoint: threshold in ImSigma*}
{\rm Im}{\Sigma^*}_{\alpha}^{\beta}(\vec{q},\omega <2\Delta)=0.
\end{equation}
Here, $\Delta$ denotes the roton (minimum) energy.

Using Eq.~(\ref{endpoint: threshold in ImSigma*}) in conjunction with
Eq.~(\ref{quasiparticles: G}) we find that the imaginary part of the
one-quasiparticle Green's function consists of a sharp peak below
and a continuum above twice the roton energy,
\begin{eqnarray}
\label{endpoint: ImG (with A)}
{\rm Im}G_{\alpha}^{\beta}(q)
        & = &-\pi A_{\alpha}^{\beta}(\vec{q})
            \delta(\omega-\omega^0_{\vec{q}}-A{\rm
            Re}\Sigma^*(q))\theta(2\Delta-\omega)\nonumber\\
        &   &+{\rm Im}M_{\alpha}^{\beta}(q),
\end{eqnarray}
where
\begin{equation}
\label{endpoint: ImM}
{\rm Im}M_{\alpha}^{\beta}(q)=\left\{
      \begin{array}{ll}
         0     &  \mbox{, $\omega<2\Delta$}\\
         \frac{A_{\alpha}^{\beta}(\vec{q})A{\rm Im}\Sigma^*(q)}
         {(\omega-\omega^0_{\vec{q}}-A{\rm Re}\Sigma^*(q))^2 +
         (A{\rm Im}\Sigma^*(q))^2}
               & \mbox{, $\omega \geq 2\Delta$}.
      \end{array}
      \right.
\end{equation}
To proceed further we use $\delta(f(x))=\sum_i
\delta(x-x_i)/f^{\prime}(x_i)$ with $f(x_i)=0$ and assume that
the equation
\begin{equation}
\label{endpoint: omega}
\omega-\omega^0_{\vec{q}}-A{\rm Re}\Sigma^*(q)=0
\end{equation}
has only one solution $\omega_{\vec{q}}$ below $2\Delta$. We then
arrive at
\begin{equation}
\label{endpoint: ImG (with Z)}
{\rm Im}G_{\alpha}^{\beta}(q)=-\pi Z_{\alpha}^{\beta}(\vec{q})
                             \delta(\omega-\omega_{\vec{q}})
                             \theta(2\Delta-\omega)+
                             {\rm Im}M_{\alpha}^{\beta}(q)
\end{equation}
with a renormalization factor given by
\begin{equation}
\label{endpoint: Z}
Z_{\alpha}^{\beta}(\vec{q})=\frac{A_{\alpha}^{\beta}(\vec{q})}
                            {|1-\frac{\partial A{\rm Re}\Sigma^*(\vec{q},\omega)}
                            {\partial \omega}|}_{\omega=\omega_{\vec{q}}}.
\end{equation}
We see that in fact, the sharp component of the one-particle
propagator can only exist below the threshold energy $2\Delta$ as
predicted by Pitaevskii. Furthermore, the only effect of
quasiparticle renormalization below the threshold is a
modification of the excitation energy and  peak strength
due to the real part of the self energy. There is no mechanism in
a model without phonons, which could change the width of the peak
below the threshold.

It is interesting to analyze the behaviour  of the peak
strength $Z_{\alpha}^{\beta}$ near the endpoint. To this end we
first need to determine some properties of the real part of the
self energy $A{\rm Re}\Sigma^*(q)$ in the vicinity of
the threshold energy. Since the real and imaginary part of the
self energy are related by
\begin{equation}
\label{endpoint: dispersion relation} A{\rm
Re}\Sigma^*(\vec{q},\omega)=-\frac{1} {\pi} {\rm P} \int {\rm
d}\epsilon \frac{A{\rm
Re}\Sigma^*(\vec{q},\omega)}{\omega-\epsilon},
\end{equation}
one can show  that the threshold in $A{\rm
Im}\Sigma^*(q)$ leads to a logarithmic  singularity in
$A{\rm Re}\Sigma^*(q)$, schematically shown in
Fig.~\ref{endpoint: singularity in aresigma}. At the threshold energy
the function is finite but its left and right derivatives are infinite;
\begin{figure}[t]
\includegraphics[height=60mm,width=70mm]{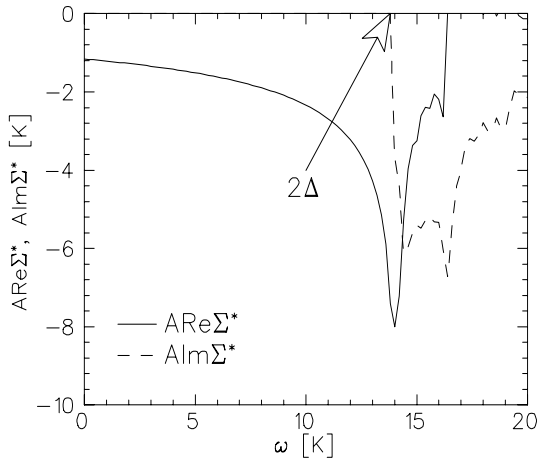}
\caption{Logarithmic singularity in $A{\rm Re}\Sigma^*(q)$ due to
the threshold in $A{\rm Im}\Sigma^*(q)$ at $\omega=2\Delta$. At
the threshold energy, the function is finite, but its left and
right derivatives go to infinity.} \label{endpoint: singularity
in aresigma}
\end{figure}
\begin{equation}
\label{endpoint: derivative of AReSigma}
|\frac{\partial A{\rm Re}\Sigma^*(\vec{q},\omega)}{\partial \omega}|
      \rightarrow \infty~~~\mbox{if~~~$\omega \rightarrow 2\Delta$}.
\end{equation}
It follows immediately from Eq.~(\ref{endpoint: derivative of
AReSigma}), that the peak strength vanishes at the endpoint
energy,
\begin{equation}
Z_{\alpha}^{\beta} (\vec{q})\rightarrow
0~~~\mbox{~~~$\omega_{\vec{q}} \rightarrow 2\Delta$}.
\end{equation}
Recall that the  dynamic susceptibility $\chi^T (q)$ shares poles
with the single-particle Green's function. Thus we can expect,
that the strength of the sharp peak in the imaginary part of
$\chi^T (q)$ (i.e. in the dynamic structure $S(q)$) vanishes at the
threshold as well. Such a behavior is in qualitative agreement
with the standard interpretation of the experimental data.

\subsection{High-momentum scattering}
\label{qualitative considerations: high-momentum scattering}

In the high-momentum region, $S(\vec{q},\omega)$ can be
increasingly well described within the impulse approximation.
In the following we will discuss the mechanism
leading to this approximation within the framework of our model. For
a more general consideration we refer to Refs.~\onlinecite{GRI93},
\onlinecite{GLY94} and \onlinecite{MAH90}.

We start by  noting that if the momentum is high enough the kinetic
energy of the quasiparticles must be very large relative to the
potential energy of their interactions. Thus interaction
effects (i.e. all terms containing ${\bf I}$ and ${\bf J}$)
can be neglected  and we obtain from Eq.~(\ref{response
function (model): chi in RPA form}) a simplified expression for
the dynamic susceptibility,
\begin{eqnarray}
\label{high-momentum scattering: chi}
\chi^T(\vec{q},\omega)  & = &n_0\sum_{\alpha\beta}g_\alpha^\beta
                             (\vec{q},\omega)+{f_0}^{21}_{21}
                             (\vec{q},\omega)+{f_0}^{21}_{12}
                             (\vec{q},\omega)\nonumber\\
                        & = &n_0\sum_{\alpha\beta}g_\alpha^\beta
                             (\vec{q},\omega)+\chi^T_R (\vec{q},\omega)
\end{eqnarray}
with
\begin{equation}
\label{high-momentum scattering: f0}
{f_0}^{\alpha \beta}_{\delta \gamma}(\vec{q},\omega)=\frac{i}{(2\pi)^4}\int {\rm d}^3 p
               \int {\rm d}\epsilon g_{\alpha}^{\delta}(\vec{p},\epsilon)
               g_{\beta}^{\gamma}(\vec{q}-\vec{p},\omega-\epsilon)
\end{equation}
and $g_{\alpha}^{\beta}$ given  by Eq.~(\ref{quasiparticles: g}).
Here, we have only two contributions to $\chi^T$. The first term
proportional to $n_0$ introduces single-quasiparticle excitations
into the density fluctuations and  produces a sharp peak in
$S(\vec{q},\omega)$. The regular part $\chi^T_R$ leads to a continuum in
$S(\vec{q},\omega)$. It is due to simultaneous excitation of two
non-interacting particles. Unlike in Eq.~(\ref{response function
(model): chi in RPA form}), the terms contributing to $\chi^T$ do
not have the same pole structure. In Ref.~\onlinecite{GRI93},
a result similar to Eq.~(\ref{high-momentum scattering: chi}),
however with different `bare' Green's functions, has been
called an improved form of the Bogoliubov approximation.

In Eq.~(\ref{high-momentum scattering: f0}), we carry out the
integration over $\epsilon$ analytically using the
theorem of residues. After that the dynamic susceptibility can be
written as
\begin{widetext}
\begin{equation}
\label{high-momentum scattering: chi1}
\chi^T(\vec{q},\omega) = n_0\sum_{\alpha\beta}g_\alpha^\beta (\vec{q},\omega)
                         +\int \frac{{\rm d}^3 p}{(2\pi)^3}
           \frac{A_2^2(\vec{p})A_1^1(\vec{q}-\vec{p})+A_1^2(\vec{p})A_2^1(\vec{q}-\vec{p})}
           {\omega-\omega_{\vec{q}-\vec{p}}^0 - \omega_{\vec{p}}^0+i\eta}.
\end{equation}
Finally, we find that
\begin{equation}
\label{high-momentum scattering: s}
S(\vec{q},\omega)  = \frac{n_0}{n}\sum_{\alpha\beta}A_{\alpha}^{\beta}(\vec{q})
                  \delta (\omega-\omega_{\vec{q}}^0)
                  +\frac{1}{n}\int \frac{{\rm d}^3 p}{(2\pi)^3}
                  [A_2^2(\vec{p})A_1^1(\vec{q}-\vec{p})+A_1^2(\vec{p})A_2^1(\vec{q}-\vec{p})]
                  \delta (\omega-\omega_{\vec{p}}^0 -\omega_{\vec{q}-\vec{p}}^0).
\end{equation}
If the dominant contribution to the integration over $\vec{p}$ in
Eq.~(\ref{high-momentum scattering: s}) is from momenta much less
than $\vec{q}$, we can use $\int {\rm d}^3 p \simeq \int_{D}{\rm
d}^3 p$, where $D=\{\vec{p}: |\vec{p}| \ll |\vec{q}|\}$. In this
region, we have $\omega_{\vec{p}}^0 +\omega_{\vec{q}-\vec{p}}^0
\simeq \omega_{\vec{q}-\vec{p}}^0$. At large enough momenta,
renormalization effects coming from the irreducible part of the
self energy ${\bf \tilde{\Sigma}}(\vec{q},\omega)$ are negligible
in comparison with the kinetic energy of the particles and we can
use the approximation
\begin{equation}
\omega_{\vec{q}-\vec{p}}^0 \simeq \frac{(\vec{q}-\vec{p})^2}{2m}
\simeq \frac{q^2}{2m}-\frac{\vec{p}\cdot \vec{q}}{m}.
\end{equation}
Here, $m$ is the free \he mass. We arrive at
\begin{equation}
\label{high-momentum scattering: s (final)}
S^{imp}(\vec{q},\omega) \simeq \frac{n_0}{n}\sum_{\alpha\beta}A_{\alpha}^{\beta}(\vec{q})
                  \delta (\omega-\frac{q^2}{2m})
+\frac{1}{n}\int \frac{{\rm d}^3 p}{(2\pi)^3}
                  [A_2^2(\vec{p})A_1^1(\vec{q}-\vec{p})+A_1^2(\vec{p})A_2^1(\vec{q}-\vec{p})]
                  \delta (\omega-\frac{q^2}{2m}+\frac{\vec{p}\cdot \vec{q}}{m}).
\end{equation}
\end{widetext}

\begin{figure}[t!]
\includegraphics[height=60mm,width=70mm]{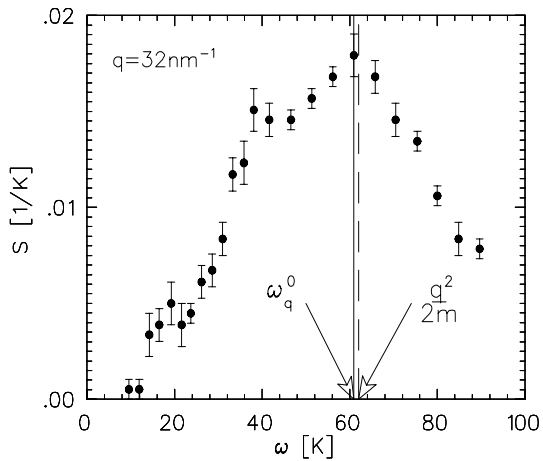}
\caption{Experimental results for $S(\vec{q},\omega)$
\cite{FAK98a} at $q=32$~nm$^{-1}$. The solid vertical line shows
the position of the peak interpreted as a quasifree peak. The
dashed line indicates the corresponding free-atom energy.}
\label{high-momentum scattering: quasifree peak}
\end{figure}

It follows from the second line  of Eq.~(\ref{high-momentum
scattering: s (final)}), that in the high momentum region
the continuum part of $S(\vec{q},\omega)$ consists of a
Doppler-broadened peak centered at the energy $q^2/2m$. In other
words, at very high momenta the continuum is dominated by
the free atom properties of the system.

Eq.~(\ref{high-momentum scattering: s (final)}) is in line  with
the Green's function reformulation of the impulse approximation
presented by Griffin~\cite{GRI93}. Some differences are due to
specific assumptions about the form of $g_{\alpha}^{\beta}$ made
within our approach.

For wavevectors $|\vec{q}|\geq 32$ nm$^{-1}$, one observes
a peak centered close to the free-atom energy (see Fig.~\ref{high-momentum
scattering: quasifree peak}). The larger the momentum
transfer the smaller the shift between the peak position and the
free-atom energy. Therefore, in \he, the impulse approximation becomes valid
already at moderately large momentum transfers. Since a small shift
in the energy of the observed `quasifree' peak indicates that the
difference between $\omega^0_{\vec{q}}$ and $q^2/2m$ is small as well,
the approximation provides a criterion for choosing a spectrum of `bare'
quasiparticles.
We will exploit this in Sec. \ref{numerical results: Landau vs. Feynman}.

\section{Preparation of the numerical analysis}
\label{technical details}

The formulas of Sec. \ref{model} provide a set of equations
enabling a numerical calculation of the dynamic susceptibility.
In this section we will briefly discuss  some details of its
solution.

\subsection{Iteration scheme}
\label{technical details: iteration scheme}

We begin by writing down the complete set of equations for the calculation
of $\chi^{\rm T}$ in a well-ordered form,
\begin{eqnarray}
\chi^T(q) & = & n_0\sum_{\alpha\beta}G_\alpha^\beta(q)+ \label{iteration scheme: chi}\\*
          &   &n_0^{1/2}\sum_{\alpha}G_\alpha^\beta(q)J_\beta^{\eta\xi}({\vec q})
                f^{21}_{\eta\xi}(q)+w^{21}_{21}(q),
                 \nonumber\\
G_{\alpha}^{\beta}(q)
          & = & \left[({\bf 1}- \frac{1}{4} {\bf gJ^{\rm T}fJ)}^{-1}\right]_\alpha^\rho
                g_\rho^\beta (q),\label{iteration scheme: G}\\
w_{21}^{21}(q)
          & = & \left[({\bf 1} - \frac{1}{4} {\bf
                fJgJ^{\rm T}})^{-1}\right]_{21}^{\xi\eta}f^{21}_{\xi\eta}(q),
                \label{iteration scheme: w}\\
f^{\delta\gamma}_{\alpha\beta}(q)
          & = & \left[({\bf 1} - \frac{1}{2}{\bf f_0I})^{-1}\right]^{\rho\sigma}_{\alpha\beta}
                ({f_0}^{\delta\gamma}_{\rho\sigma}+{f_0}^{\delta\gamma}_{\sigma\rho}),
                \label{iteration scheme: f}\\
{f_0}^{\delta\gamma}_{\alpha\beta}(q)
          & = & \frac{i}{(2\pi)^4}\int {\rm d}^4 p G_{\alpha}^{\delta} (-p+\frac{q}{2})
                G_{\beta}^{\gamma}(p+\frac{q}{2})
\label{iteration scheme: f0}
\end{eqnarray}
We see from the above equations that the functions
${f_0}^{\delta\gamma}_{\alpha\beta}(q)$ are all we need in
order to calculate the dynamic susceptibility $\chi^{\rm T}(q)$.
However, determination of these functions is not a trivial task,
since they are contained implicitly on the right hand side of
Eq.~(\ref{iteration scheme: f0}). We will solve these
integral equations self-consistently via the following iteration
procedure: We start with $G_\alpha^\beta=g_\alpha^\beta$ (i.e.
${\Sigma^*}_\alpha^\beta (q)=0$), and calculate ${f_0}_{\alpha
\beta}^{\delta \gamma}$. The result is used to evaluate
$f_{\alpha \beta}^{\delta \gamma}$ from which new
$G_\alpha^\beta$ are obtained. With these $G_\alpha^\beta$ we again
calculate ${f_0}_{\alpha \beta}^{\delta \gamma}$ and the
calculation is repeated until self-consistency is established.
The procedure is illustrated schematically in the diagram in Fig.~\ref{iteration
scheme: diagram}. Note that technically the iteration cycle has
similarities with a calculation made by G\"{o}tze and
L\"{u}cke~\cite{GOE76} who considered density excitations in
superfluid \he using Mori's formalism.
\begin{figure}[t!]
\includegraphics{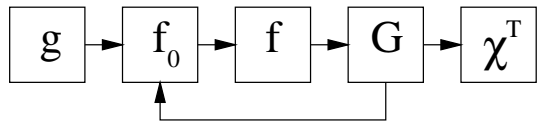}
\caption{Iteration scheme. See text for more details.}
\label{iteration scheme: diagram}
\end{figure}

Obviously, we need the `bare' quasiparticle
propagators $g_\alpha^\beta (q)$ (i.e. $\omega_{\vec{q}}^0$ and
the strengths $A_{\alpha} ^{\beta}(\vec{q})$) and the
interactions ${\bf I}(\vec{q})$ and ${\bf J}(\vec{q})$ in order to
calculate $\chi^T$. For the
sake of simplicity, we will treat them as
independent parameters of the model. Neglecting the connection
between the input quantities has some consequences. For instance,
if one changes only the `bare' spectrum $\omega^0_{\vec{q}}$ and keeps the
interactions fixed, one arrives effectively at a different
physical system. In other words, the result of the iteration
depends on the `bare' spectrum. That is the reason why we need some
criteria like high-momentum scattering (see Sec. \ref{qualitative
considerations: high-momentum scattering}) in order to choose a
suitable input spectrum for our calculation.

We will use the resulting  one-quasiparticle spectrum $\omega_{\vec{q}}$,
which is given by Eq.~(\ref{endpoint: omega}), as a consistency check: a
solution is consistent, if after some number of iteration steps the
renormalized spectrum does not change any more.

\subsection{Calculation of ${f_0}_{\alpha \beta}^{\delta \gamma}(q)$}
\label{technical details: calculation of f0}

In order to facilitate our iteration scheme we want to carry out
the integration over energy in Eq.~(\ref{iteration scheme: f0})
analy\-tically.

According to Eqs.~(\ref{quasiparticles: G}) and  (\ref{endpoint:
threshold in ImSigma*}) we can split ${f_0}_{\alpha
\beta}^{\delta \gamma}(q)$ into three terms:
\begin{equation}
{f_0}_{\alpha \beta}^{\delta \gamma}(q)=f_0^{1,1}(q)+f_0^{1,m}(q)+f_0^{m,m}(q).
\label{calculation of f0: three terms in f0}
\end{equation}
For the sake of simplicity we have suppressed the greek indices
on the right hand side of the above expression. We find (recall
that $p=(\vec{p},\epsilon)$ and $q=(\vec{q},\omega)$)

\begin{widetext}
\begin{equation}
f_0^{1,1}(q)  = i\int \frac{{\rm d}^3p}{(2\pi)^3}\int \frac{{\rm d}\epsilon}{2\pi}
                  \frac{A_{\alpha}^{\delta}(\vec{p})\Theta(2\Delta-\epsilon)}
                  {\epsilon-\omega_{\vec{p}}^0-A{\rm Re}\Sigma^*(p)+i\eta}\;
                  \frac{A_{\beta}^{\gamma}(\vec{q}-\vec{p})\Theta(2\Delta-\omega+\epsilon)}
                  {\omega-\epsilon-\omega_{\vec{q}-\vec{p}}^0-A{\rm
                  Re}\Sigma^*(q-p)+i\eta^{\prime}}
\label{calculation of f0: f011}
\end{equation}
\end{widetext}
for the single-quasiparticle-single-quasiparticle term,
\begin{eqnarray}
f_0^{1,m}(q)  & = & i\int \!\! \frac{{\rm d}^3p}{(2\pi)^3}\int \!\!
                   \frac{{\rm
                  d}\epsilon}{2\pi} \left[ \frac{A_{\alpha}^{\delta}(\vec{p})
                  M_{\beta}^{\gamma}(q-p)\Theta(2\Delta-\epsilon)}
                  {\epsilon-\omega_{\vec{p}}^0-A{\rm Re}\Sigma^*(p)+i\eta} \right.
                  \nonumber\\
              &   & \left. +\frac{M_{\alpha}^{\delta}(p)A_{\beta}^{\gamma}(\vec{q}-\vec{p})
                  \Theta(2\Delta-\omega+\epsilon)}
                  {\omega-\epsilon-\omega_{\vec{q}-\vec{p}}^0-A{\rm
                  Re}\Sigma^*(q-p)+i\eta^{\prime}} \right]
\label{calculation of f0: f01m}
\end{eqnarray}
for the single-quasiparticle-continuum contribution and finally
\begin{equation}
f_0^{m,m}(q) = i\int \frac{{\rm d}^3p}{(2\pi)^3}\int \frac{{\rm
               d}\epsilon}{2\pi}M_{\alpha}^{\beta}(p)M_{\beta}^{\gamma}(q-p)
\label{calculation of f0: f0mm}
\end{equation}
for the continuum-continuum part coming from the product of two
single-quasiparticle propagators. Here, we have introduced the
auxiliary function
\begin{equation}
M_{\alpha}^{\beta}(q)=G_{\alpha}^{\beta}(q)\Theta(\omega-2\Delta).
\label{calculation of f0: M}
\end{equation}
Note, that the imaginary part of $M_{\alpha}^{\beta}$ is given by
Eq.~(\ref{endpoint: ImM}).

In order to carry out the energy integrals in Eqs.~(\ref{calculation
of f0: f011})-(\ref{calculation of f0: f0mm}) we first make an
analytic continuation of the integrands into the region of
imaginary energies and then perform a contour integration. We are
dealing with retarded propagators, so the  contour should run
along the real energy axis and be closed by a large arc in the
lower half plane. Since the Green's functions behaves like
$\omega^{-1}$ for $|\omega|\rightarrow \infty$, Jordan's lemma
ensures that the contribution from the arc at infinity vanishes
and we can evaluate the integral along the real axis by using of
the theorem of residues.

Apart from $f_{0}^{m,m}(q)$, which is analytical in the whole
energy plane, all other terms in ${f_0}_{\alpha \beta}^{\delta
\gamma}$ have  single poles in the lower half plane. After
calculating the corresponding residues we get for the imaginary
part of ${f_0}_{\alpha \beta}^{\delta \gamma}$
\begin{widetext}
\begin{eqnarray}
{\rm Im}{f_0}_{\alpha \beta}^{\delta \gamma}(q)
              & = &-\pi \int \frac{{\rm d}^3 p}{(2\pi)^3}\{ Z_{\alpha}^{\delta}(\vec{p})
                   Z_{\beta}^{\gamma}(\vec{q}-\vec{p})\delta(\omega-\omega_{\vec{p}}-
                   \omega_{\vec{q}-\vec{p}})\Theta(2\Delta-\omega_{\vec{p}})
                   \Theta(2\Delta-(\omega-\omega_{\vec{p}}))\nonumber\\*
              &   &+ Z_{\alpha}^{\beta}(\vec{p}){\rm
                    Im}M_{\beta}^{\gamma}(\vec{q}-\vec{p},
                           \omega-\omega_{\vec{p}})\Theta(2\Delta-\omega_{\vec{p}})
                          +{\rm Im}M_{\alpha}^{\delta}(\vec{p},\omega-\omega_{\vec{q}-\vec{p}})
                          Z_{\beta}^{\gamma}(\vec{q}-\vec{p})
                          \Theta(2\Delta-\omega_{\vec{q}-\vec{p}})
                          \}
\label{calculation of f0: Imf0}
\end{eqnarray}
\end{widetext}
with $Z_{\alpha}^{\delta}$ and $\omega_{\vec{p}}$ defined in Sec.
\ref{qualitative considerations: endpoint}. Obviously, the
continuum-continuum contribution to ${\rm Im}{f_0}_{\alpha
\beta}^{\delta \gamma}$ vanishes after carrying out the
energy integrals.

We could also calculate the real part of ${f_0}_{\alpha
\beta}^{\delta \gamma}$ in an analogous way. However, since the
function is analytic off the real axis, its real and imaginary
parts are connected  through the causality relation
\begin{equation}
{\rm Re}{f_0}_{\alpha \beta}^{\delta \gamma}(\vec{q},\omega) = -\frac{1}{\pi}
{\rm P}\int {\rm d}\epsilon \frac{{\rm Im}{f_0}_{\alpha \beta}^{\delta \gamma}
(\vec{q},\epsilon)}{\omega-\epsilon},
\label{calculation of f0: dispersion relation}
\end{equation}
where ${\rm P}$ denotes a Cauchy principal value. Since a calculation
of the expression~(\ref{calculation of f0: dispersion relation}) requires
less numerical effort than the
direct method based on Eqs.~(\ref{calculation of f0:
f011})-(\ref{calculation of f0: f0mm}), we will use it in our
numerics.

\section{Numerical results}
\label{numerical results}

Our calculation scheme contains several parameters:  the
condensate density $n_0$, the matrix functions ${\bf I}$, ${\bf
P}$, and ${\bf A}$, and the free quasiparticle spectrum
$\omega_{\vec{q}}^0$. For the condensate fraction we take
$n_0=0.1n$, where $n$ is the density of the liquid. This
assumption is in agreement with  the condensate density
extracted from experimental data~\cite{SNO92} and supported by
Path Integral Monte Carlo calculations \cite{CEP86,CEP87}.

In a recent  paper \cite{SZW00}, we have presented first
numerical results for $S(\vec{q},\omega)$ obtained in a
simplified version of the calculation with a momentum independent self energy
of the quasiparticles and constant interactions between them. Since
such an approximation led to some inconsistencies in the calculated
spectra, in the present work we have removed this restriction for the
self energy. However, for the sake of simplicity, we still treat the
interaction functions as momentum and index independent, i.e.
\begin{equation}
I_{\alpha \beta}^{\delta \gamma}(\vec{q})=g_4,~~~J_{\alpha \beta}^{\gamma}(\vec{q})=g_3.
\end{equation}
Here, $g_4$ and $g_3$ are phenomenological constants. Similarly,
we assume that the renormalization functions
$A_{\alpha}^{\beta}(\vec{q})$'s are momentum independent and equal
in all channels (i.e. $A_{\alpha}^ {\beta}(\vec{q})=A$ for
$\alpha , \beta =1,2$). Obviously, this is not in line with the
standard choices  made to make a contact with ZRS-like models
from the microscopic theory (see discussion  in Sec. \ref{model:
quasiparticles}). However, under these assumptions,  the elements
of the Beliaev-Green's matrix function are identical. If we now
carry out summations over greek indices in our expressions, we
are able to obtain a scheme of ZRS type.

In each iteration step, we have evaluated the dynamic  structure
factor for 56 $|\vec{q}|$-values and 501 $\omega$-values. As
mentioned in Sec. \ref{technical details: iteration scheme}, we
have used the resulting spectrum $\omega_{\vec{q}}$ as a consistency check: a
solution  was consistent, if after a certain number of iteration
steps the renormalized spectrum did not change any more.
Depending on the input parameters 10-50 steps were necessary to
obtain a consistent solution.
\begin{figure}[b]
\includegraphics[height=60mm,width=70mm]{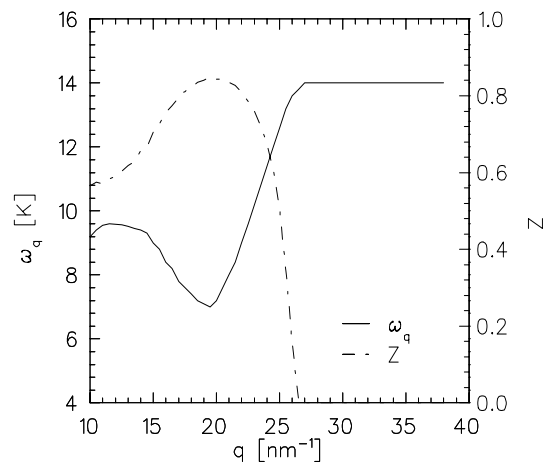}
\caption{The renormalized excitation spectrum as calculated from
Eq.~(\ref{endpoint: omega}) (solid curve) and the
resulting strength $Z(\vec{q})$ of the sharp peak in ${\rm
Im}G(q)$ (dashed-dotted line).}
\label{pitaevskii: spectrum and strength}
\end{figure}

\subsection{Pitaevskii singularities}
\label{numerical results: Pitaevskii singularities}

Let us first consider some general features of the results within
a specific example with the Landau spectrum as the `bare' quasiparticle
spectrum $\omega_{\vec{q}}^0$.

\begin{figure}[t]
\includegraphics[height=60mm,width=70mm]{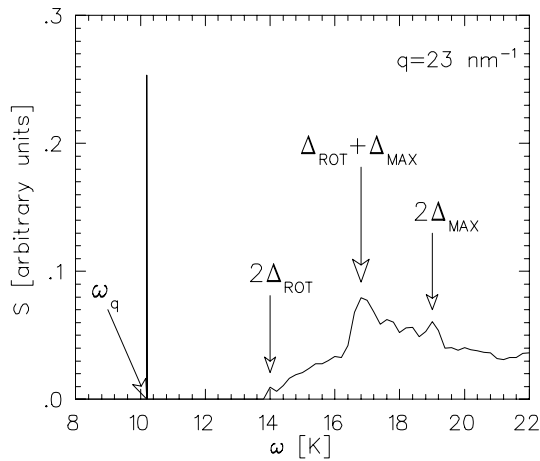}
\caption{\dsf at $q=23$ nm$^{-1}$ calculated using the Landau free quasiparticle
spectrum. The structure observed in the continuum is partly due to the roton-roton, roton-maxon and maxon-maxon pair excitations. They correspond to the van Hove singularities in the density of states function.}
\label{pitaevskii: S(q,omega)}
\end{figure}

Results for the renormalized single-quasiparticle spectrum $\omega_{\vec{q}}$,
which is defined by Eq.~(\ref{endpoint: omega}), are shown in
Fig.~(\ref{pitaevskii: spectrum and strength}). $\omega_{\vec{q}}$
is characterized by an maximum at the energy $\Delta_M=9.6$K, a
minimum at $\Delta_R=7.0$K and an endpoint at $(2\vec{q}_R,2\Delta_R)$,
where $\vec{q}_R$ is the wavevector of the roton minimum. At higher momentum
transfers the renormalized spectrum differ qualitatively from the `bare' spectrum.
It bends toward $2\Delta_R$ due to quasiparticle decay.  Since the
density of states is high near the local extrema of the spectrum,
we expect to observe a structure in the continuum part of \dsf, whose origin is
associated with the roton-roton, roton-maxon and maxon-maxon pair excitations.
They are as well due to quasiparticle decay as discussed by Pitaevskii~\cite{PIT59}.
Naively, one could expect to also observe  features stemming from the plateau
near the endpoint. But excitations associated to this part of the
dispersion curve have a vanishing weight  $Z_{\alpha}^{\beta}$ (the
dashed-dotted line in Fig.~\ref{pitaevskii: spectrum and strength})
and corresponding singularities will not appear in the pair excitation
spectrum.

In Fig.~\ref{pitaevskii: S(q,omega)}, we show results of the same
calculation for \dsf at $q=23$ nm$^{-1}$. The dynamic structure factor
consists of a sharp peak lying at the renormalized single-quasiparticle
energy $\omega_{\vec{q}}$ and a continuum, which is not featureless
at all. As expected, we observe peaks at energies $2\Delta_R$,
$\Delta_R+\Delta_M$ and $2\Delta_M$, respectively. Moreover, there is
structure, which we cannot assign to particular regions of the single-quasiparticle
excitation spectrum. We believe this to be a numerical defect. However,
it has not been eliminated, since it is not important for the following
discussion.

As mentioned above, we do not take phonons into account in our calculation.
This results in a gap between the sharp peak and the continuum in \dsf.
For the same reason, there is no mechanism in the model, which would change
the width of the sharp peak below the decay threshold.

\subsection{Landau vs. Feynman spectrum}
\label{numerical results: Landau vs. Feynman}

\begin{figure*}[t!]
\includegraphics[height=60mm,width=150mm]{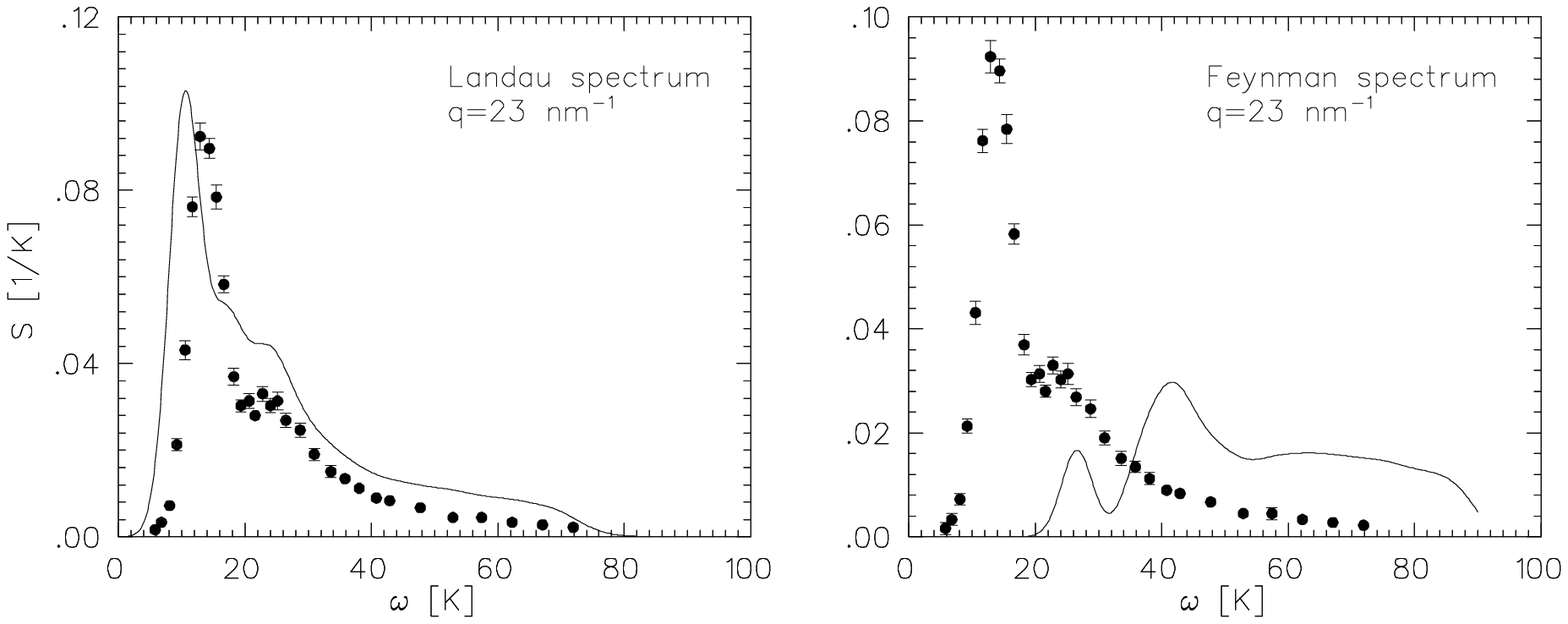}
\caption{$S(\vec{q},\omega)$  at $q=23$~nm$^{-1}$ calculated with
different `bare' spectra: the Landau spectrum (left figure) and
the Feynman one (right figure). Interactions $g_3$ and $g_4$ and
renormalization function $A$ are the same in both cases:
$g_3=0.5$~Knm$^{3/2}$, $g_4=-0.04$~Knm$^3$ and
$A=0.25$.} \label{landau vs. feynman: lanvsfey1}
\end{figure*}

\begin{figure*}[t]
\includegraphics[height=60mm,width=150mm]{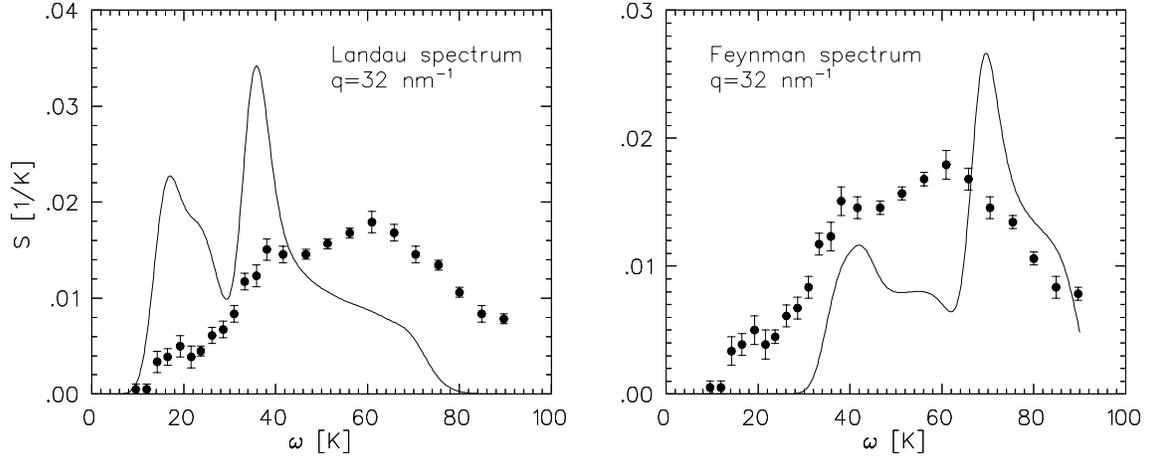}
\caption{$S(\vec{q},\omega)$ at $q=32$~nm$^{-1}$. See caption of
Fig.~\ref{landau vs. feynman: lanvsfey1} for more details.}
\label{landau vs. feynman: lanvsfey2}
\end{figure*}

In this section, we want to compare results  for
$S(\vec{q},\omega)$, which differ only in the `bare' spectrum
$\omega_{\vec{q}}^0$ from each other (Fig.~(\ref{landau vs. feynman:
lanvsfey1}) and (\ref{landau vs. feynman: lanvsfey2})).
All other input parameters are the same. They were chosen
to reproduce the observed dynamic structure factor at
$q=23$ nm$^{-1}$ in the case of the Landau `bare' quasiparticle
spectrum as well as possible. The values are as follows:
\begin{equation}
g_3=0.5~{\rm Knm}^{3/2},~~g_4=-0.04~{\rm Knm}^3,~~A=0.25.
\label{landau vs. feynman: other parameters}
\end{equation}
To include experimental resolution we  have convoluted the
calculated $S(\vec{q},\omega)$ with a Gaussian function. The
standard deviation of the Gaussian was taken to be $2.46$~K in
agreement with the  experimental resolution obtained by F{\aa}k and Bossy
\cite{FAK98a}.

In Fig.~\ref{landau vs. feynman: lanvsfey1},   results for
$S(\vec{q},\omega)$ at $q=23$ nm$^{-1}$ are shown in comparison
to experiment. We see that, indeed, in the case of the Landau
`bare' spectrum (left figure), the calculated results describe
the two peak structure present in experimental data well.
The lower energy peak corresponds to single-quasiparticle excitations.
The higher energy structure  comes from the Pitaevskii singularities
discussed in the previous section. They have merged into a broad
peak due to the convolution with a Gaussian. The theoretical and
experimental strengths of the peaks agree to a reasonable
accuracy. However, the calculated peaks appear at somewhat
different energies as compared to the experimental data.

The results in case of the Feynman spectrum are much worse. We
still observe a two peak structure, but the peaks are shifted to
higher energies. Moreover, the strength of the
single-quasiparticle  peak is too small in relation to that of
the second one. The shift in the energies indicates that the
input parameters were too small to force the spectrum down
to the experimental one. Actually, one can try to find some
parameters, which lead to better results with the Feynman
spectrum. But the stronger the vertex functions, the smaller the
difference between the potential and kinetic energy of
quasiparticles and the worse the approximation about locality of
the vertices (\ref{response function (model): interaction}).

Fig.~\ref{landau vs. feynman: lanvsfey1} suggests that the Landau
spectrum is  a better choice for our calculation. However, as we know
from Sec. \ref{qualitative considerations: high-momentum scattering},
it cannot describe the quasifree peak, which becomes dominant for
$q\geq 32$~nm$^{-1}$, very well. The results at $q=32$~nm$^{-1}$ are
presented in Fig.~\ref{landau vs. feynman: lanvsfey2}. In case of the Landau
spectrum (left figure), we observe two peaks in the calculated
$S(\vec{q},\omega)$. The first one is due to Pitaevskii's
singularites. Its position agrees with the experimental one to a
good accuracy, however its  strength does not fit the measured
one any more. We will come back to this feature later. The
continuum is dominated by a peak centered at $\omega\approx 35$~K.
It turns out, that this is the Landau energy of an excitation
corresponding to the wavevector $q=32$~nm$^{-1}$ and the peak can
be interpreted as quasifree peak. However, it appears at energy
$25$~K smaller than the observed one.

In case of the Feynman  spectrum we still observe the shift of the
whole structure to higher energies. But as expected, now the
quasifree peak is qualitatively better
described. Note, that now the shift is not the same for both
peaks. The position of the quasifree peak is determined by the
`bare' spectrum, while the first peak (due to  Pitaevskii
singularities) comes from the renormalized one. So, its position
should depend on other input parameters, i.e. vertex and
renormalization functions.

As we see, neither the Landau  spectrum nor the Feynman spectrum
appears to be  a good candidate for the `bare' quasiparticle
dispersion relation, if one wants to describe
$S(\vec{q},\omega)$ in a wide region of momentum transfers. While
the Landau spectrum is better in the kinematic region, where the
Pitaevskii's singularities play the dominant role, the Feynman
spectrum is more suited for the quasifree region. Since our goal
is a description of both regions, we decided
to modify the `bare' spectrum. In the following we  use a
phenomenological spectrum given by the dashed-dotted curve in
Fig.~\ref{landau vs. feynman: modified landau spectrum}. The
phonon-maxon-roton part of the new $\omega_{\vec{q}}^0$ agrees
with the Landau spectrum (solid line), but it approaches the
Feynman energy (dashed line) much faster above the roton minimum.
\begin{figure}[t]
\includegraphics[height=60mm,width=70mm]{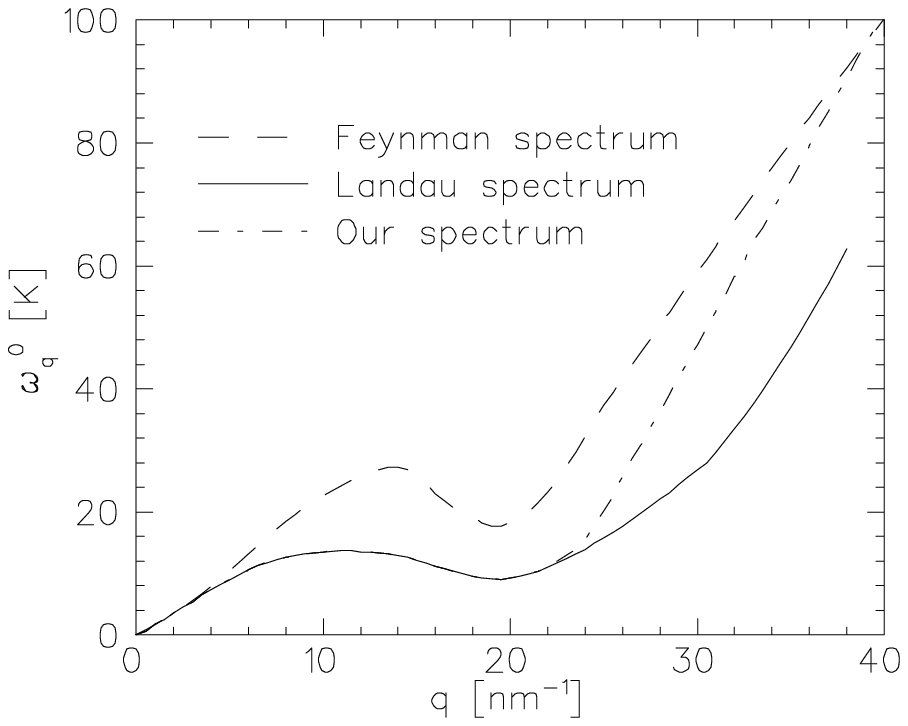}
\caption{Our  `free' quasiparticle spectrum (dotted-dashed line):
we modify the Landau curve (solid line) in the region above the
roton minimum such that it approaches Feynman spectrum (dashed
line) much faster.} \label{landau vs. feynman: modified landau
spectrum}
\end{figure}
\begin{figure}[!tb]
\includegraphics[height=150mm,width=60mm]{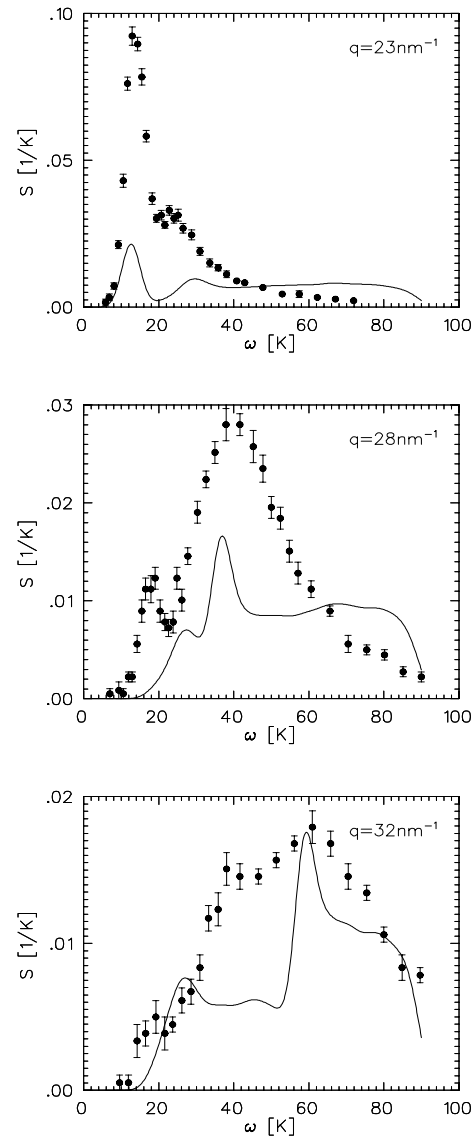}
\caption{$S(\vec{q},\omega)$  calculated from the
modified Landau spectrum as `bare' quasiparticle spectrum. The
input parameters were chosen to give a reasonable fit to the experimental
data at $q=32$~nm$^{-1}$: $g_3=0.3$~Knm$^{3/2}$, $g_4=0.19$~Knm$^3$ and $A=0.4$.}
\label{S(q,omega): S fitted at 32 nm**-1}
\end{figure}
\begin{figure}[!tb]
\includegraphics[height=150mm,width=60mm]{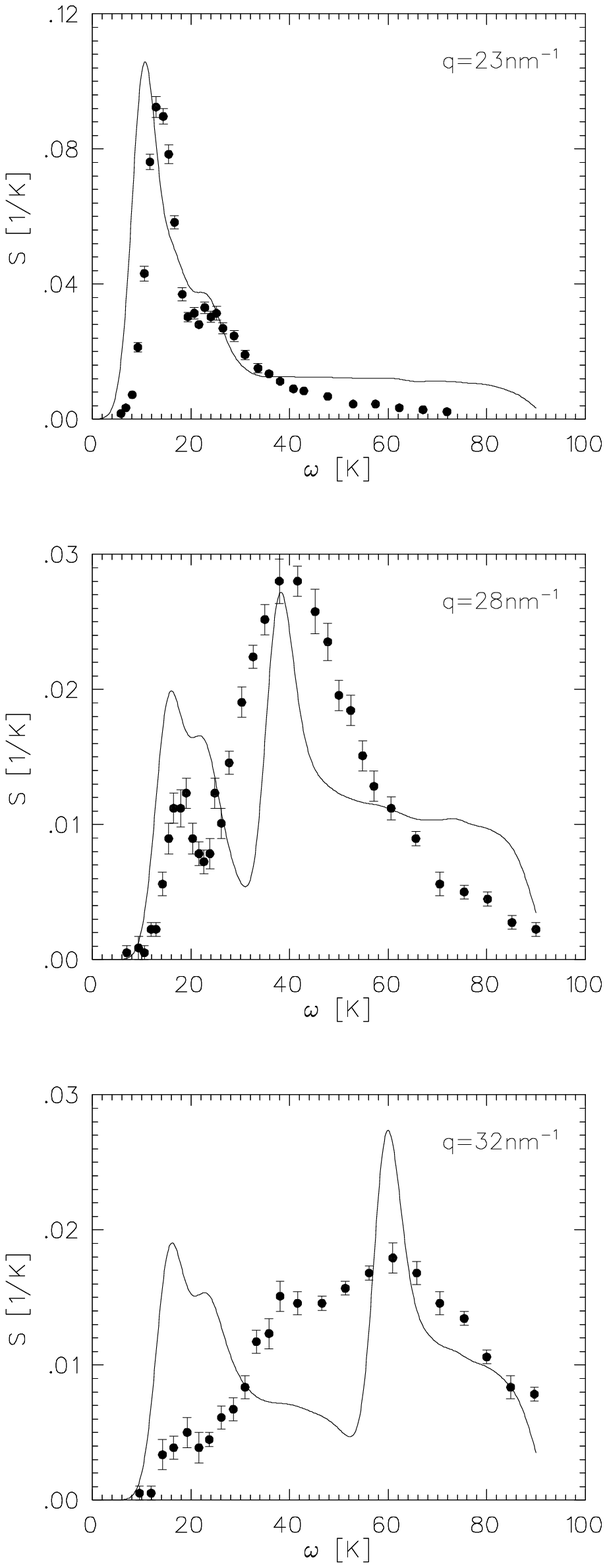}
\caption{$S(\vec{q},\omega)$ calculated from the
modified Landau spectrum, but with different interaction
parameters:  $g_3=0.5$~Knm$^{3/2}$, $g_4=-0.08$~Knm$^3$ and
$A=0.265$. They were chosen to give a reasonable fit to the experimental
data at $q=23$~nm$^{-1}$.}
\label{S(q,omega): S fitted at 23 and 28 nm**-1}
\end{figure}

Apart from  the Pitaevskii's singularities and the quasi\-free
peak, in case of the Feynman spectrum there is an indication for
a third peak lying between them (Fig.~\ref{landau vs. feynman: lanvsfey2}).
It corresponds to the third experimental peak observed at
$\omega\approx 39$K. But its strength is much to small compared with
experiment. We will discuss this issue in more detail below.

\subsection{$S(\vec{q},\omega)$ with `new' quasiparticles}
\label{numerical results: S(q,omega)}

The effects of the modified `bare' quasiparticle spectrum on \dsf are shown
in Fig.~\ref{S(q,omega): S fitted at 32 nm**-1}. In this calculation, we
used
\begin{equation}
g_3=0.3~{\rm Knm}^{3/2},~~g_4=0.19~{\rm Knm^3},~~A=0.4
\label{S(q,omega): other parameters at 32nm**-1}
\end{equation}
and the Gaussian resolution  function discussed in the previous
section. As expected, the results at $q=32$~nm$^{-1}$ are now much
better than in the case of the Landau spectrum. First of all, we
see that the quasifree peak is sitting at the right position.
Moreover, weights of the dominant peaks agree to a better accuracy
with experiment. There is also  evidence for the third peak
observed in experiment. But its intensity is still much to small.
However, the results at $q=23$~nm$^{-1}$ and $q=28$~nm$^{-1}$ are
not satisfactory. In both cases the qualitative structure of \dsf
is still reasonable, but the agreement with the experiment is now
much worse than at $q=32$~nm$^{-1}$. From the analysis of the
results we are led to the conclusion, that it is difficult to
describe \dsf  in a wide momentum region in the framework of
ZRS-like models with momentum independent input parameters. We
will remove this restriction consistently during further
refinement of the calculation.

If the above  conclusion was  true, we should be able to find
another set of parameters, which leads to better results for \dsf
at $q=23$~nm$^{-1}$ or $q=28$~nm$^{-1}$. Indeed, we have found such
parameters, which lead to reasonable results in the region
of smaller momenta. In Fig.~\ref{S(q,omega): S fitted at 23 and
28 nm**-1}, we present $S(\vec{q},\omega)$ calculated with
\begin{equation}
g_3=0.5~{\rm Knm}^{3/2},~~g_4=-0.08~{\rm Knm}^3~\mbox{and}~A=0.265.
\label{S(q,omega): other parameters at 23nm**-1}
\end{equation}
Now, \dsf at $q=23$~nm$^{-1}$  is described quite well. However,
if we compare Fig.~\ref{S(q,omega): S fitted at 23 and 28 nm**-1}
with the results of the previous section obtained with the Landau
spectrum, we see, that the modification of the spectrum results
in an unreasonable plateau at higher energies. This again points to
a possible momentum dependence of the interaction parameters. Although
somewhat worse than at $q=23$~nm$^{-1}$, the calculated spectrum
at $q=28$~nm$^{-1}$ is still reasonable compared to the experiment
in this special case of input parameters. And the results at
$q=32$~nm$^{-1}$, which are now worse than in the previous case,
seem to confirm our conclusion that input parameters must be momentum
dependent.

From the input parameters reported in (\ref{S(q,omega): other parameters
at 32nm**-1}) and (\ref{S(q,omega): other parameters at 23nm**-1}) an
interesting feature of the irreducible four-point vertex function
$g_4$ follows. We have achieved the best fit to the data at intermediate
momenta for a negative $g_4$. But at high momenta, $g_4$ has to be
positive in order to describe $S(\vec{q},\omega)$ properly.
Thus, the effective interaction between
quasiparticles changes its character from an attraction at intermediate
momenta to a repulsion at higher ones. We will come back to this feature
in forthcoming work~\cite{SZW01}.

As we know from Eq.~(\ref{response function (model): chi}), there
are three contri\-butions to $S(\vec{q},\omega)$, which come from
the single-quasi\-particle, interference and two-quasiparticle
terms. We will refer to them as $S^{(1)}$, $S^{(int)}$ and
$S^{(2)}$, respectively. The energy integrated strengths of these
contributions are reported in Table \ref{S(q,omega): partial
contributions}. There are two important features shown in the
table. First, $S^{(int)}$ reduces the strength of the sharp
single-quasiparticle peak at $q=23$~nm$^{-1}$ and of the continuum
at $q=32$~nm$^{-1}$. Secondly, and more important, in all cases
the structure factor is dominated by the two-particle contribution
$S^{(2)}$. Thus the expression
\begin{equation}
S(\vec{q},\omega) \propto {\rm Im}G(\vec{q},\omega),
\label{S(q,omega): S propto ImG}
\end{equation}
which was used in the literature (e.g. Ref.~\onlinecite{HAS74})
in order to calculate \dsf within ZSR-type models, cannot describe the
data in the momentum region under consideration, since the most
relevant term is neglected.
\begin{table}[t]
\centering
\begin{tabular}{|c|c|c||c||c|}
  \hline
           & \multicolumn{2}{c||}{$q=23$nm$^{-1}$}
                   & $q=28$nm$^{-1}$
                           &$q=32$nm$^{-1}$
                             \\ \cline{2-5}
           & peak  & continuum & continuum &continuum
                             \\ \hline
$S^{(1)}$  & 0.07  & 0.03      & 0.10  & 0.15    \\ \hline
$S^{(int)}$&-0.23  & 0.23      & 0.01  & -0.11   \\ \hline
$S^{(2)}$  &0.77   & 1.03      & 0.87  & 0.57 \\ \hline
\end{tabular}
\caption{Energy integrated  contributions to $S(\vec{q},\omega)$
from the single-quasiparticle, interference and two-quasiparticle
terms at different momentum transfers. The contributions at
$q=23$~nm$^{-1}$ and $q=28$~nm$^{-1}$ have been obtained with
$g_3=0.5$~Knm$^{3/2}$, $g_4=-0.08$~Knm$^3$ and $A=0.265$. For
$q=32$~nm$^{-1}$, we used $g_3=0.3$~Knm$^{3/2}$,
$g_4=0.19$~Knm$^3$ and $A=0.4$. The dynamic structure factor at
$q=28$~nm$^{-1}$ and $q=32$~nm$^{-1}$ contains only a structured
continuum. At these momenta, there is no sharp
peak in \dsf, since the strength $Z_{\alpha}^{\beta}$ given by
Eq.~(\ref{endpoint: Z}) vanishes.} \label{S(q,omega): partial
contributions}
\end{table}

\subsection{Comparisons with other calculations}
\label{numerical results: comparisons}

In this section, we want to  compare our results with those
obtained by other methods. We will limit ourselves to the work of
G\"{o}tze and L\"{u}cke \cite{GOE76} and of Manousakis and
Phandaripande \cite{MAN84,MAN86}. All theoretical results are compared
to the experimental data of F{\aa}k and Bossy \cite{FAK98a}.

\begin{figure*}[t!]
\includegraphics[height=5cm]{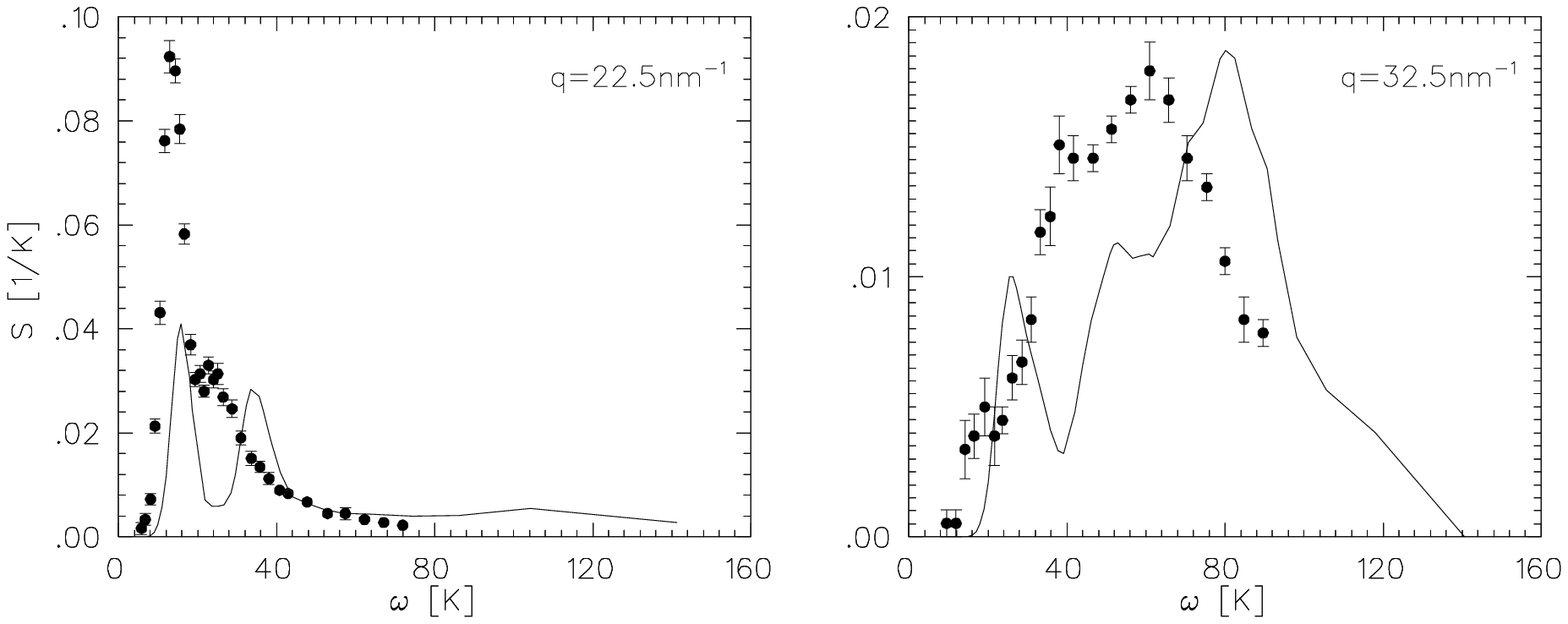}
\caption{Calculation of \dsf by G\"{o}tze and
L\"{u}cke~\cite{GOE76} using the memory function formalism at two
selected momenta compared to experiment. The calculated spectra
have been convoluted with a Gaussian resolution function as
discussed in Sec.
 \ref{numerical results: Landau vs. Feynman}.}
\label{comparisons: goetze-luecke}
\vspace{0.5cm}
\includegraphics[height=5cm]{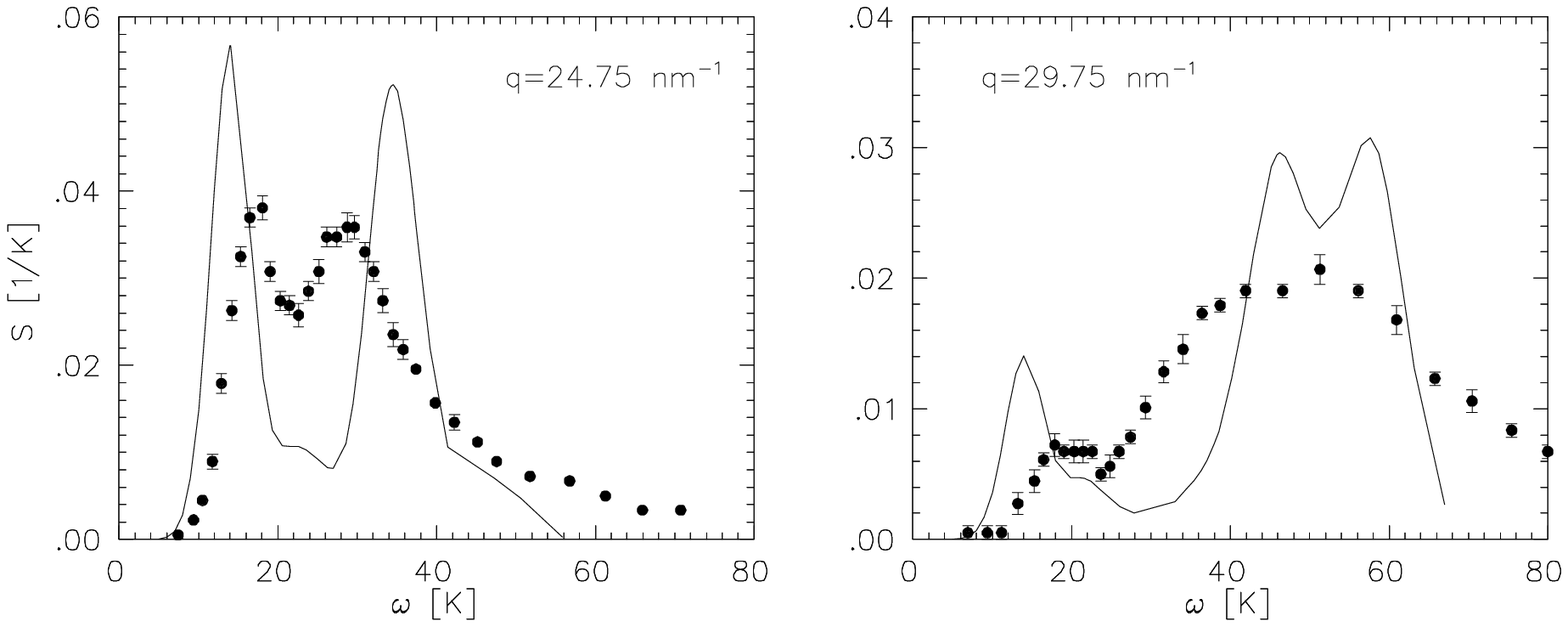}
\caption{Calculation of \dsf by Manousakis and Phandaripande~\cite{MAN84,MAN86}
using correlated basis function methods. See caption of
Fig.~\ref{comparisons: goetze-luecke} for more details.}
\label{comparisons: manousakis-phandaripande}
\vspace{0.5cm}
\includegraphics[height=5cm]{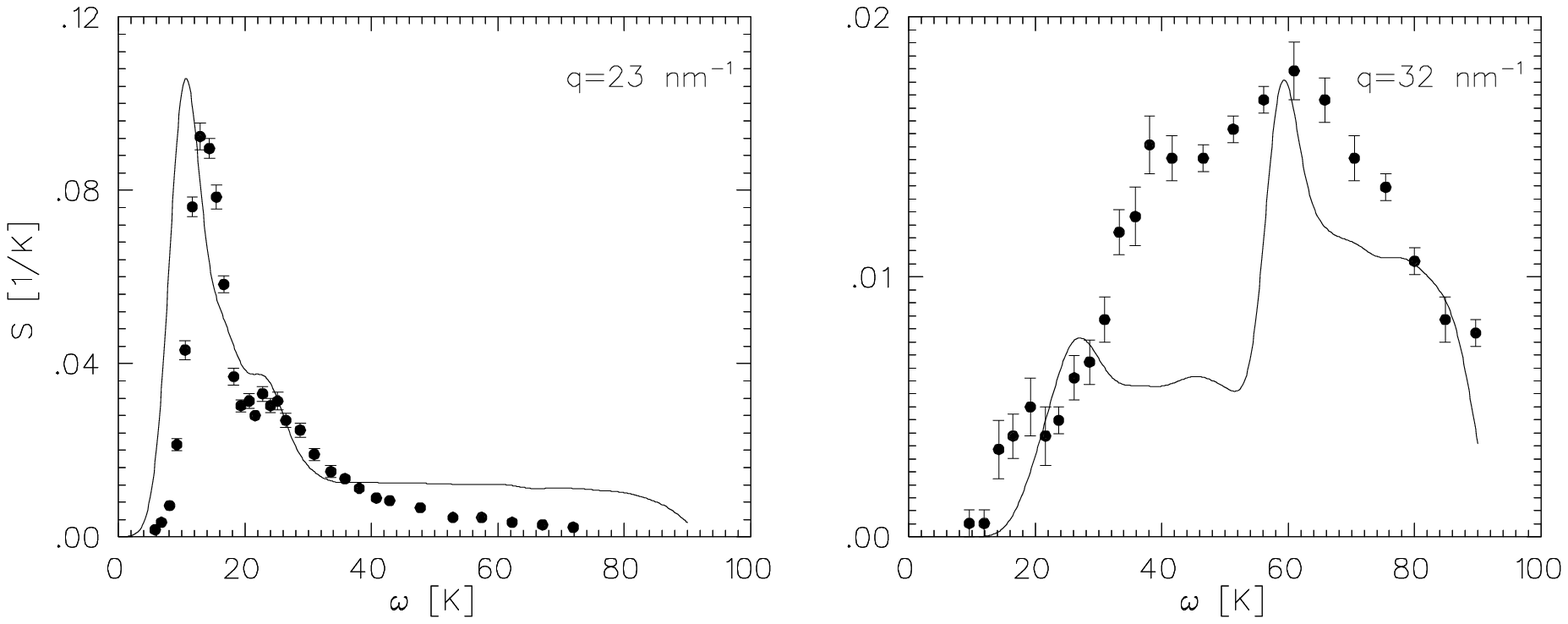}
\caption{Our best results for \dsf at two selected momenta.  The case $q=23$~nm$^{-1}$
corresponds to $g_3=0.5$~Knm$^{3/2}$, $g_4=-0.08$~Knm$^3$ and $A=0.265$. For
$q=32$~nm$^{-1}$, we have used $g_3=0.3$~Knm$^{3/2}$, $g_4=0.19$~Knm$^3$ and $A=0.4$.}
\label{comparisons: our best results}
\end{figure*}

G\"{o}tze  and L\"{u}cke presented a detailed analysis of
$S(\vec{q},\omega)$ for superfluid \he at $T=0$ within
the memory function formalism. Their results at two selected
momenta  are presented in Fig.~\ref{comparisons: goetze-luecke}.
In order to make comparisons with experiment we have convoluted
the calculated spectra with the Gaussian resolution function
discussed in Sec \ref{numerical results: Landau vs. Feynman}.

In Fig.~\ref{comparisons: manousakis-phandaripande}, we show
$S(\vec{q},\omega)$ calculated by Manousakis and Pandharipande,
who used correlated-basis-function methods  at $T=0$.
Again, the theoretical spectra are convoluted with the above mentioned
Gaussian resolution function.

In order to facilitate comparison of the different methods our best
results for the dynamic structure factor are plotted
again in Fig.~\ref{comparisons: our best results}. Recall, that
each plot corresponds to different parameter sets in order to simulate
an effective momentum dependence of our parameters.

It follows  from the Figs.~\ref{comparisons:
goetze-luecke}-\ref{comparisons: our best results}  that  all
three methods are significantly missing strength in the energy region
between 30 - 50~K at momenta around 30~nm$^{-1}$. This fact may indicate
that there is a lack of understanding of the underlying physics.

\section{Conclusions}
\label{conclusions}

In this paper, we presented a calculation of the dynamic structure factor
of superfluid \he in the intermediate and high momentum transfer region within a model based
on the Gavoret-Nozi\`{e}res~\cite{GAV64} microscopic theory. After the introduction of
quasiparticles, we obtained an RPA-like expression for the
density-density correlation function in a model, which has similarities
to the phenomenological field theory of
Zawadowski, Ruvalds and Solana~\cite{ZAW72}, but treats the condensate explicitly.
We evaluated $S(\vec{q},\omega)$ numerically for the special case of
momentum independent interactions between the quasiparticles. Our results suggest
the following conclusions:
\begin{enumerate}
\def\theenumi{\roman{enumi}}
\def\labelenumi{(\roman{enumi})}
\item The quasiparticles, that were introduced here, can be interpreted as helium atoms
      renormalized by the two-particle irreducible part of
      the self energy. Quasifree scattering may be used to determine
      the high energy part of the `bare' quasiparticle spectrum
      appropriately.
\item \dsf consists of three terms: one- and
      two-quasiparticle excitations and a interference term, $S^{(1)}$, $S^{(2)}$
      and $S^{(int)}$, respectively. All terms have the same pole structure and
      appear to be equally important in the momentum region at and above the roton
      minimum. Thus the expression $S(\vec{q},\omega)\propto {\rm Im}
      G(\vec{q},\omega)$ often used in literature in order to calculate the
      dynamic structure factor in ZRS-type models neglects important terms.
      In the absence of the condensate only the $S^{(2)}$ term remains.
\item A model which employs momentum independent interactions
      cannot account quantitatively for the neutron scattering data in
      a wide momentum region. \label{conlusions: iii}
\item The two-quasiparticle interaction $g_4$ should be attractive at
      intermediate momenta  and repulsive at higher momentum transfers.
\item As is seen from the comparison with other calculations, our model
      provides an alternative description of the experimental data.
      However, no method describes qualitatively the data at
      $|\vec{q}|\geq 30$~nm$^{-1}$. This may indicate a lack of understanding of
      the underlying physics.
\end{enumerate}

We now turn to a somewhat more detailed discussion of the
point~(\ref{conlusions: iii}). At present, we do not know very
well $g_3$ and $g_4$ as functions of $\vec{q}$. Thus for the sake
of simplicity we assumed, that they are weakly momentum dependent
and can be approximated by constants. The same holds for the
residues $A_\alpha^\beta$. For the chosen parameters, we evaluated
\dsf at 56 values of $|\vec{q}|$. Our results  show that it is
difficult to describe \dsf in a wide momentum transfer region
within a model with constant parameters $g_3$, $g_4$ and $A$. To
avoid this problem, we have `simulated' a momentum dependence by
calculating the structure factor with different parameter sets.
For each set, we picked out from the 56 spectra those with
reasonably good agreement with experiment. In this way, we got
sets of parameters, that led to good results in different momentum
intervals.

Clearly, we have made progress
towards the development of a quasiparticle model, which treats condensate
and non-condensate terms in \dsf on the same footing. However, the numerical
calculation needs further refinement  which will be addressed in future work.

\begin{acknowledgments}
We would like to thank Professors M.~L\"{u}cke and W.~Weller for useful
discussions, Dr.~W.~Apel for reading the manuscript and Dr.~R.~Scherm for
discussions about the experimental data and his continued support.
\end{acknowledgments}

\bibliography{he4}

%
%

%
%

\end{document}